\documentclass[preprint,11pt,a4paper]{elsarticle}
\usepackage{mathtools}
\usepackage{amsfonts}
\usepackage{graphicx}
\usepackage{amsmath}
\usepackage{amssymb}
\usepackage{bm}
\usepackage{color}
\usepackage{import}
\usepackage{hyperref}
\usepackage{appendix}
\usepackage{afterpage}

\biboptions{authoryear,sort}
\journal{Medical Image Analysis}

\def\Snospace~{}

\newcommand{\modified}[1]{#1}

\DeclareMathOperator*{\argmin}{arg\,min}
\DeclareMathOperator*{\argmax}{arg\,max}

\DeclareMathOperator*{\diag}{diag}
\renewcommand{\vec}[1]{\mathbf{#1}}

\newcommand{\CompModel}{\ensuremath{f}}

\newcommand{\MDP}{\ensuremath{\mathcal{M}}}

\newcommand{\State}{\ensuremath{s}}
\newcommand{\StateSet}{\ensuremath{\mathcal{S}}}
\newcommand{\nStates}{{\ensuremath{n_{\StateSet}}}}
\newcommand{\StateOpt}{\ensuremath{\hat{\State}}}

\newcommand{\Action}{\ensuremath{a}}
\newcommand{\ActionSet}{\ensuremath{\mathcal{A}}}
\newcommand{\nActions}{{\ensuremath{n_{\ActionSet}}}}

\newcommand{\Transitions}{\ensuremath{\mathcal{T}}}
\newcommand{\Rewards}{\ensuremath{\mathcal{R}}}
\newcommand{\Discount}{\ensuremath{\gamma}}

\newcommand{\Policy}{\ensuremath{\pi}}
\newcommand{\PolicyOpt}{\ensuremath{\Policy^*}}

\newcommand{\PolicyStochastic}{\ensuremath{\tilde{\pi}}}
\newcommand{\PolicyStochasticOpt}{\ensuremath{\PolicyStochastic^*}}

\newcommand{\Param}{\ensuremath{x}}
\newcommand{\ParamVec}{\ensuremath{\vec{\Param}}}
\newcommand{\nParams}{\ensuremath{{n_{\ParamVec}}}}

\newcommand{\ParamDomain}{\ensuremath{\Omega}}

\newcommand{\ModelState}{\ensuremath{y}}
\newcommand{\ModelStateVec}{\ensuremath{\vec{\ModelState}}}
\newcommand{\nModelStates}{\ensuremath{{n_{\ModelStateVec}}}}

\newcommand{\Measurement}{\ensuremath{z}}
\newcommand{\MeasurementVec}{\ensuremath{\vec{\Measurement}}}

\newcommand{\Objective}{\ensuremath{c}}
\newcommand{\ObjectiveVec}{\ensuremath{\vec{\Objective}}}
\newcommand{\nObjectives}{\ensuremath{{n_{\ObjectiveVec}}}}
\newcommand{\ObjectiveSet}{\ensuremath{\mathcal{C}}}

\newcommand{\Episode}{\ensuremath{e}}
\newcommand{\EpisodeSet}{\ensuremath{\mathcal{E}}}

\newcommand{\nEpisodeSteps}{\ensuremath{{n_{\Episode\text{-steps}}}}}

\newcommand{\Centroid}{\ensuremath{\xi}}
\newcommand{\CentroidVec}{\ensuremath{\vec{\Centroid}}}

\newcommand{\Convergence}{\ensuremath{\psi}}
\newcommand{\ConvergenceVec}{\ensuremath{\boldsymbol{\Convergence}}}

\newcommand{\ReferenceValue}{\ensuremath{\delta}}
\newcommand{\ReferenceValueVec}{\ensuremath{\boldsymbol{\ReferenceValue}}}

\newcommand{\StateMapping}{\ensuremath{\phi}}

\newcommand{\nTrainingSamples}{\ensuremath{{n_{\text{samples}}}}}
\newcommand{\nDatasets}{\ensuremath{{n_{\text{datasets}}}}}
\begin{document}
\begin{frontmatter}
%
%
%
\title{A Self-Taught Artificial Agent for Multi-Physics Computational Model Personalization}
%
%
%
\author[siemensDE,fau]{Dominik~Neumann\corref{cor1}}
\ead{dominik.neumann@siemens.com}
\author[siemensUS]{Tommaso~Mansi}
\author[siemensRO,uniRO]{Lucian~Itu}
\author[siemensUS]{Bogdan~Georgescu}
\author[heidelberg]{Elham~Kayvanpour}
\author[heidelberg]{Farbod~Sedaghat-Hamedani}
\author[heidelberg]{Ali~Amr}
\author[heidelberg]{Jan~Haas}
\author[heidelberg]{Hugo~Katus}
\author[heidelberg]{Benjamin~Meder}
\author[fau]{Stefan~Steidl}
\author[fau]{Joachim~Hornegger}
\author[siemensUS]{Dorin~Comaniciu}
\address[siemensDE]{Medical Imaging Technologies, Siemens Healthcare GmbH, Erlangen, Germany}
\address[siemensUS]{Medical Imaging Technologies, Siemens Healthcare, Princeton, USA}
\address[fau]{Pattern Recognition Lab, FAU Erlangen-N\"urnberg, Erlangen, Germany}
\address[siemensRO]{Siemens Corporate Technology, Siemens SRL, Brasov, Romania}
\address[uniRO]{Transilvania University of Brasov, Brasov, Romania}
\address[heidelberg]{Department of Internal Medicine III, University Hospital Heidelberg, Germany}
\cortext[cor1]{Corresponding author}
%
%
%
%
\begin{abstract}
Personalization is the process of fitting a model to patient data, a critical step towards application of multi-physics computational models in clinical practice.
Designing robust personalization algorithms is often a tedious, time-consuming, model- and data-specific process.
We propose to use artificial intelligence concepts to learn this task, inspired by how human experts manually perform it.
The problem is reformulated in terms of reinforcement learning.
In an off-line phase, Vito, our self-taught artificial agent, learns a representative decision process model through exploration of the computational model: it learns how the model behaves under change of parameters.
The agent then automatically learns an optimal strategy for on-line personalization.
The algorithm is model-independent; applying it to a new model requires only adjusting few hyper-parameters of the agent and defining the observations to match.
The full knowledge of the model itself is not required.
Vito was tested in a synthetic scenario, showing that it could learn how to optimize cost functions generically.
Then Vito was applied to the inverse problem of cardiac electrophysiology and the personalization of a whole-body circulation model.
The obtained results suggested that Vito could achieve equivalent, if not better goodness of fit than standard methods, while being more robust (up to 11\% higher success rates) and with faster (up to seven times) convergence rate.
Our artificial intelligence approach could thus make personalization algorithms generalizable and self-adaptable to any patient and any model.
\end{abstract}
%
%
%
\end{frontmatter}
%
%
%
\section{Introduction}
\label{sec:intro}
%
Computational modeling attracted significant attention in cardiac research over the last decades \citep{frangi2001three,noble2002modeling,hunter2003integration,kerckhoffs2008cardiac,clayton2011models,kuijpers2012modeling,krishnamurthy2013patient}.
It is believed that computational models can improve patient stratification and therapy planning.
They could become the enabling tool for predicting disease course and therapy outcome, ultimately leading to improved clinical management of patients suffering from cardiomyopathies \citep{kayvanpour2015towards}.
A crucial prerequisite for achieving these goals is precise model personalization: the computational model under consideration needs to be fitted to each patient. 
However, the high complexity of cardiac models and the often noisy and sparse clinical data still hinder this task.

A wide variety of manual and (semi-)automatic model parameter estimation approaches have been explored, including \cite{augenstein2005method,schmid2006myocardial,wang2009modelling,sermesant2009personalised,aguado2010computational,konukoglu2011efficient,aguado2011patient,delingette2012personalization,chabiniok2012estimation,xi2013estimation,marchesseau2013personalization,le2013current,prakosa2013cardiac,wallman2014computational,zettinig2014data,neumann2014robust,neumann2014automatic,itu2014parestimation,seegerer2015estimation,wong2015velocity}.
Most methods aim to iteratively reduce the misfit between model output and measurements using optimization algorithms, for instance variational \citep{delingette2012personalization} or filtering \citep{marchesseau2013personalization} approaches.
Applied blindly, those techniques could easily fail on unseen data, if not supervised, due to parameter ambiguity, data noise and local minima \citep{neumann2014robust,wallman2014computational,konukoglu2011efficient}.
Therefore, complex algorithms have been designed combining cascades of optimizers in a very specific way to achieve high levels of robustness, even on larger populations, i.e.~10 or more patients \citep{kayvanpour2015towards,seegerer2015estimation,neumann2014automatic}.
However, those methods are often designed from tedious, trial-and-error-driven manual tuning, they are model-specific rather than generic, and their generalization to varying data quality cannot be guaranteed.
On the contrary, if the personalization task is assigned to an experienced human, given enough time, he almost always succeeds in manually personalizing a model for any subject (although solution uniqueness is not guaranteed, but this is inherent to the problem).

There are several potential reasons why a human expert is often superior to standard automatic methods in terms of personalization accuracy and success rates.
First, an expert is likely to have an intuition of the model's behavior from his prior knowledge of the physiology of the modeled organ.
Second, knowledge about model design and assumptions, and model limitations and implementation details certainly provide useful hints on the ``mechanics'' of the model.
Third, past personalization of other datasets allows the expert to build up experience.
The combination of prior knowledge, intuition and experience enables to solve the personalization task more effectively, even on unseen data.

Inspired by humans and contrary to previous works, we propose to address the personalization problem from an artificial intelligence (AI) perspective.
In particular, we apply reinforcement learning (RL) methods \citep{sutton1998reinforcement} developed in the AI community to solve the parameter estimation task for computational physiological models.
With its roots in control theory on the one hand, and neuroscience theories of learning on the other hand, RL encompasses a set of approaches to make an artificial agent learn from experience generated by interacting with its environment.
Contrary to standard (supervised) machine learning \citep{bishop2006pattern}, where the objective is to compute a direct mapping from input features to a classification label or regression output, RL aims to learn \emph{how to} perform tasks.
The goal of RL is to compute an optimal problem-solving strategy (agent behavior), e.g.~a strategy to play the game ``tic-tac-toe'' successfully.
In the AI field, such a behavior is often represented as a policy, a mapping from states, describing the current ``situation'' the agent finds itself in (e.g.~the current locations of all ``X'' and ``O'' on the tic-tac-toe grid), to actions, which allow the agent to interact with the environment (e.g.~place ``X'' on an empty cell) and thus influence that situation.
The key underlying principle of RL is that of reward \citep{kaelbling1996reinforcement}, which provides an objective means for the agent to judge the outcome of its actions.
In tic-tac-toe, the agent receives a high, positive reward if the latest action led to a horizontal, vertical or diagonal row full of ``X'' marks (winning), and a negative reward (punishment) if the latest action would allow the opponent to win in his next move.
Based on such rewards, the artificial agent learns an optimal winning policy through trial-and-error interactions with the environment.

RL was first applied to game \citep[e.g.][]{tesauro1994td} or simple control tasks.
However, the past few years saw tremendous breakthroughs in RL for more complex, real-world problems~\citep[e.g.][]{nguyen2011model,kveton2012kernel,barreto2014practical}.
Some noteworthy examples include \cite{mulling2013learning}, where the control entity of a robot arm learned to select appropriate motor primitives to play table tennis, and \cite{mnih2015human}, where the authors combine RL with deep learning to train an agent to play 49 Atari games, yielding better performance than an expert in the majority of them.

\begin{figure}[t]
  \centering
  \includegraphics[width=.8\columnwidth]{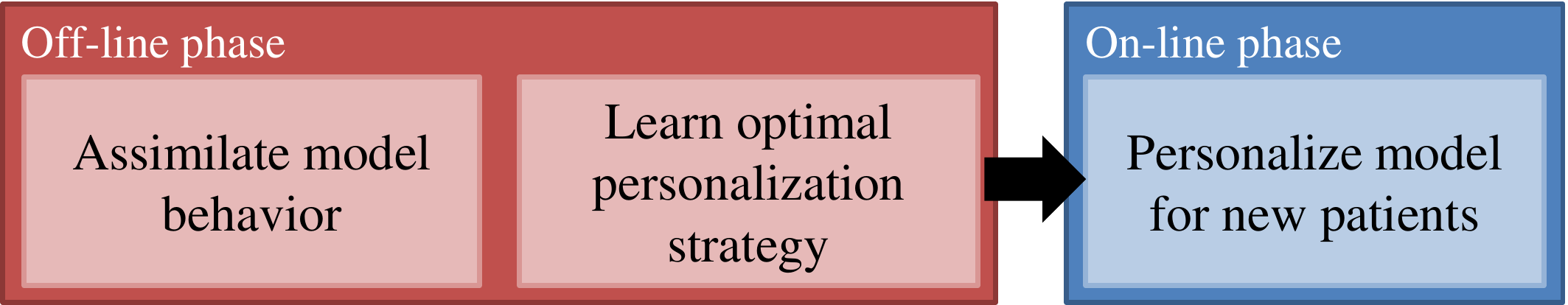}
  \caption{Overview of Vito: a self-taught artificial agent for computational model personalization, inspired by how human operators approach the personalization problem.}
  \label{fig:intro:general_overview}
\end{figure}

Motivated by these recent successes and building on our previous work \citep{neumann2015vito}, we propose an RL-based personalization approach, henceforth called \textit{Vito}, with the goal of designing a framework that can, for the first time to our knowledge, learn by itself how to estimate model parameters from clinical data while being model-independent.
As illustrated in \autoref{fig:intro:general_overview}, first, like a human expert, Vito assimilates the behavior of the physiological model under consideration in an off-line, one-time only, data-driven exploration phase.
From this knowledge, Vito learns the optimal strategy using RL~\citep{sutton1998reinforcement}.
The goal of Vito during the on-line personalization phase is then to sequentially choose actions that maximize future rewards, and therefore bring Vito to the state representing the solution of the personalization problem.
To setup the algorithm, the user needs to define what observations need to be matched, the allowed actions, and a single hyper-parameter related to the desired granularity of the state-space.
Then everything is learned automatically.
The algorithm does not depend on the underlying model.

Vito was evaluated on three different tasks.
First, in a synthetic experiment, convergence properties of the algorithm were analyzed.
Then, two tasks involving real clinical data were evaluated: the inverse problem of cardiac electrophysiology and the personalization of a lumped-parameter model of whole-body circulation.
The obtained results suggested that Vito can achieve equivalent (or better) goodness of fit as standard optimization methods, increased robustness and faster convergence rates.

\modified{A number of novelties and improvements over \cite{neumann2015vito} are featured in this manuscript.
First, an automatic, data-driven state-space quantization method is introduced that replaces the previous manual technique.
Second, the need to provide user-defined initial parameter values is eliminated by employing a new data-driven technique to initialize personalization of unseen data.
Third, a stochastic personalization policy is introduced, for which the previously used standard deterministic policy is a special case.
Fourth, the convergence properties are evaluated in parameter space using a synthetic personalization scenario.
In addition, thorough evaluation of Vito's performance with increasing amount of training samples was conducted and personalization of the whole-body circulation model was extended to several variants involving two to six parameters.
Finally, the patient database used for experimentation was extended from 28 to 83 patients for the cardiac electrophysiology experiments, and from 27 to 56 for the whole-body circulation experiments.}

The remainder of this manuscript is organized as follows.
\autoref{sec:method} presents the method.
In \autoref{sec:experiments}, the experiments are described and the results are presented.
\autoref{sec:conclusion} concludes the manuscript with a summary and discussions about potential limitations and extensions of the method.
%
%
%
\section{Method}
\label{sec:method}
%
This section presents the reinforcement-learning (RL) framework for computational model personalization.
\autoref{subsec:method:mdp} introduces Markov decision process (MDP).
\autoref{subsec:method:perso2mdp} defines the personalization problem and how it can be reformulated in terms of an MDP.
\autoref{subsec:method:exploration} describes how the artificial agent, Vito, learns how the model behaves. 
Next, \autoref{subsec:method:state_quantization} provides details about state-space quantization, and \autoref{subsec:method:transitions} describes how the model knowledge is encoded in the form of transition probabilities.
All steps mentioned so far are performed in an off-line training phase.
Finally, \autoref{subsec:method:execution} explains how the learned knowledge is applied on-line to personalize unseen data.
%
%
\subsection{Model-based Reinforcement Learning}
\label{subsec:method:mdp}
%
%
\subsubsection{MDP Definition}
%
A crucial prerequisite for applying RL is that the problem of interest, here personalization, can be modeled as a Markov decision process (MDP).
An MDP is a mathematical framework for modeling decision making when the decision outcome is partly random and partly controlled by a decision maker \citep{sutton1998reinforcement}.
Formally, an MDP is a tuple $\MDP = ( \StateSet, \ActionSet, \Transitions, \Rewards, \Discount )$, where:
\begin{itemize}
  \item $\StateSet$ is the finite set of states that describe the agent's environment, \modified{$\nStates$ is the number of states}, and $\State_t \in \StateSet$ is the state at time $t$.
  \item $\ActionSet$ is the finite set of actions, which allow the agent to interact with the environment, \modified{$\nActions$ is the number of actions}, and $\Action_t \in \ActionSet$ denotes the action performed at time $t$.
  \item $\Transitions: \StateSet \times \ActionSet  \times \StateSet \rightarrow [0;1]$ is the stochastic transition function, where $\Transitions(\State_t, \Action_t, \State_{t+1})$ describes the probability of arriving in state $\State_{t+1}$ after the agent performed action $\Action_t$ in state $\State_t$.
  \item $\Rewards: \StateSet \times \ActionSet \times \StateSet \rightarrow \mathbb{R}$ is the scalar reward function, where $r_{t+1} = \Rewards(\State_t, \Action_t, \State_{t+1})$ is the immediate reward the agent receives at time $t+1$ after performing action $\Action_t$ in state $\State_t$ resulting in state $\State_{t+1}$.
  \item $\Discount \in [0;1]$ is the discount factor that controls the importance of future versus immediate rewards.
\end{itemize}
%
\subsubsection{Value Iteration}
%
The value of a state, $V^*(\State)$, is the expected discounted reward the agent accumulates when it starts in state \State~and acts optimally in each step:
\begin{equation}
  V^*(\State) =
    E
    \left\{
      \sum_{k=0}^\infty
        \Discount^k r_{t+k+1} \middle| \State_t = \State
    \right\}
  \enspace ,
  \label{eq:method:stateVal}
\end{equation}
where $E\{\}$ denotes the expected value given the agent always selects the optimal action, and $t$ is any time step. \modified{Note that the discount factor $\Discount$ is a constant and the superscript $k$ its exponent.}
$V^*$ can be computed using value iteration \citep{sutton1998reinforcement}, an iterative algorithm based on dynamic programming. 
In the first iteration $i = 0$, let $V_i: \StateSet \rightarrow \mathbb{R}$ denote an initial guess for the value function that maps states to arbitrary values.
Further, let $Q_i: \StateSet \times \ActionSet \rightarrow \mathbb{R}$ denote the $i$\textsuperscript{th} ``state-action value function''-guess, which is computed as:
\begin{equation}
  Q_i(\State, \Action) = \sum_{\State' \in \StateSet} \Transitions(\State,\Action,\State') \left[ \Rewards(\State, \Action, \State') + \Discount V_i(\State') \right] \enspace .
  \label{eq:method:Q}
\end{equation}
Value iteration iteratively updates $V_{i+1}$ from the previous $Q_i$: 
\begin{equation}
  \forall \State \in \StateSet: \quad V_{i+1}(\State) = \max_{\Action \in \ActionSet} Q_i(\State, \Action) \enspace ,
  \label{eq:method:bellman}
\end{equation}
until the left- and right-hand side of \autoref{eq:method:bellman} are equal for all $\State \in \StateSet$; then $V^* \leftarrow V_{i+1}$ and $Q^* \leftarrow Q_{i+1}$.
From this equality relation, also known as the Bellman equation \citep{bellman1957dynamic}, one can obtain an optimal problem-solving strategy for the problem described by the MDP (assuming that all components of the MDP are known precisely).
It is encoded in terms of a deterministic optimal policy $\PolicyOpt: \StateSet \rightarrow \ActionSet$: 
\begin{equation}
  \PolicyOpt(\State) = \argmax_{\Action \in \ActionSet} Q^*(\State, \Action)\enspace ,
  \label{eq:method:optimal_policy}
\end{equation}
i.e.~a mapping that tells the agent in each state the optimal action to take.
%
\subsubsection{Stochastic Policy}
%
In this work not all components of the MDP are known precisely, instead some are approximated from training data.
Value iteration, however, assumes an exact MDP to guarantee optimality of the computed policy.
Therefore, instead of relying on the deterministic policy \PolicyOpt~(\autoref{eq:method:optimal_policy}), a generalization to stochastic policies \PolicyStochasticOpt~is proposed here to mitigate potential issues due to approximations.
Contrary to \autoref{eq:method:optimal_policy}, where for each state only the one action with maximum $Q^*$-value is considered, a stochastic policy stores several candidate actions with similar high $Q^*$-value and returns one of them through a random process each time it is queried.
To this end, the $Q^*(\State,\cdot)$-values for a given state \State~are first normalized:
\begin{equation}
  \tilde{Q}_\State^*(\Action) =
  \frac
  {
    Q^*(\State,\Action) - \min_{\Action' \in \ActionSet} [Q^*(\State,\Action')]
  }
  {
    \max_{\Action' \in \ActionSet} [Q^*(\State,\Action')] - \min_{\Action' \in \ActionSet} [Q^*(\State,\Action')]
  }
  \enspace .
  \label{eq:method:stochastic_policy_probs}
\end{equation}
All actions whose normalized $\tilde{Q}_\State^*$-value is below a threshold of $\epsilon = \frac{4}{5}$ (set empirically and used throughout the entire manuscript) are discarded, while actions with large values are stored as potential candidates.
Each time the stochastic policy is queried, $\Action = \PolicyStochasticOpt_\epsilon(\State)$, it returns one of the candidate actions \Action~selected randomly with probability proportional to its $\tilde{Q}_\State^*$-value: $\tilde{Q}_\State^*(\Action) / \sum_{\Action'} \tilde{Q}_\State^*(\Action')$; the sum is over all candidate actions $\Action'$.
%
%
%
\subsection{Reformulation of the Model Personalization Problem into an MDP}
\label{subsec:method:perso2mdp}
%
\subsubsection{Problem Definition}
\begin{figure}[t]
  \centering
  \includegraphics[width=.8\columnwidth]{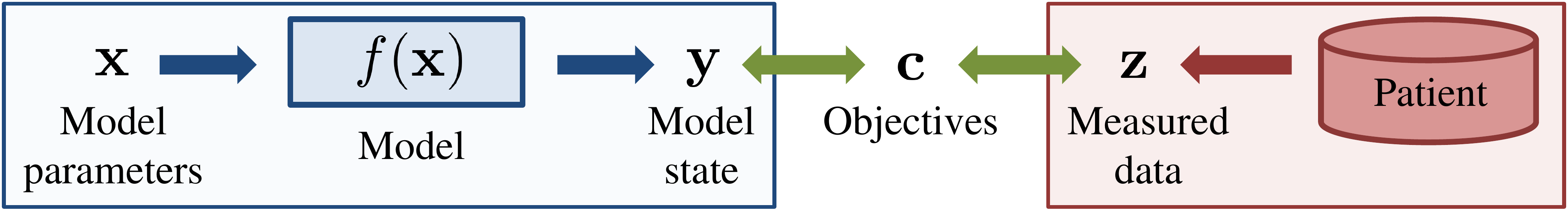}
  \caption{A computational model \CompModel~is a dynamic system that maps model input parameters \ParamVec~to model state (output) variables \ModelStateVec. The goal of personalization is to tune \ParamVec~such that the objectives \ObjectiveVec, defined as the misfit between \ModelStateVec~and the corresponding measured data \MeasurementVec~of a given patient, are optimized (the misfit is minimized).}
  \label{fig:method:model_diagram}
\end{figure}

%
As illustrated in \autoref{fig:method:model_diagram}, any computational model \CompModel~is governed by a set of parameters $\ParamVec = ( \Param_1, \dots, \Param_\nParams )^\top$, \modified{where $\nParams$ denotes the number of parameters}.
\ParamVec~is bounded within a physiologically plausible domain $\ParamDomain$, and characterized by \modified{a number of $\nModelStates$} (observable) state variables $\ModelStateVec = ( \ModelState_1, \dots, \ModelState_\nModelStates )^\top$.
The state variables can be used to estimate $\ParamVec$.
Note that some parameters may be pre-estimated or assigned fixed values.
The goal of personalization is to optimize a set of \modified{$\nObjectives$} objectives $\ObjectiveVec = ( \Objective_1, \dots, \Objective_\nObjectives )^\top$.
The objectives are scalars defined as $\Objective_i = d(\ModelState_i, \Measurement_i)$, where $d$ is a measure of misfit, and $\Measurement_i$ denotes the patient's measured data (\MeasurementVec) corresponding to $\ModelState_i$.
In this work $d(\ModelState_i, \Measurement_i) = \ModelState_i - \Measurement_i$.
Personalization is considered successful if all user-defined convergence criteria $\ConvergenceVec = ( \Convergence_1, \dots, \Convergence_{\nObjectives} )^\top$ are met.
The criteria are defined in terms of maximum acceptable misfit per objective: $\forall i \in \{1, \dots, \nObjectives\}: |\Objective_i| < \Convergence_i$.
%
%
\subsubsection{Problem Reformulation}
\label{subsubsec:method:reformulation}
%
Personalization is mapped to a Markov decision process as follows:

\noindent\textbf{States}: An MDP state encodes the misfit between the computed model state (outcome of forward model run) and the patient's measurements.
Thus, MDP states carry the same type of information as objective vectors \ObjectiveVec, yet the number of MDP states has to be finite (\autoref{subsec:method:mdp}), while there are an infinite number of different objective vectors due to their continuous nature.
Therefore the space of objective vectors in $\mathbb{R}^\nObjectives$ is reduced to a finite set of \emph{representative states}: the MDP states \StateSet, each $\State \in \StateSet$ covering a small region in that space.
One of those states, $\StateOpt \in \StateSet$, encodes personalization success as it is designed such that it covers exactly the region where all convergence criteria are satisfied.
The goal of Vito is to learn how to reach that state.

\noindent\textbf{Actions}: Vito's actions modify the parameters $\ParamVec$~to fulfill the objectives $\ObjectiveVec$.
An action $\Action \in \ActionSet$ consists in either in- or decrementing one parameter $\Param_i$ by $1\times$, $10\times$ or $100\times$ a user-specified reference value $\ReferenceValue_i$ with $\ReferenceValueVec = (\ReferenceValue_1, \dots, \ReferenceValue_\nParams)^\top$.
This empirically defined quantization of the intrinsically continuous action space yielded good results for the problems considered in this work. 

\noindent\textbf{Transition function}: \Transitions~encodes the agent’s knowledge about the computational model \CompModel~and is learned automatically as described in \autoref{subsec:method:transitions}.

\noindent\textbf{Rewards}: Inspired by the ``mountain car'' benchmark \citep{sutton1998reinforcement}, the rewards are defined as always being equal to $\Rewards(\State, \Action, \State') = -1$ (punishment), except when the agent performs an action resulting in personalization success, i.e.~when $\State' = \StateOpt$.
In that case, $\Rewards(\cdot, \cdot, \StateOpt) = 0$ (no punishment).

\noindent\textbf{Discount factor}: The large discount factor $\Discount = 0.99$ encourages policies that \modified{favor future over immediate rewards, as Vito should always prefer the long-term goal of successful personalization to short-term appealing actions in order to reduce the risk of local minima.}
%
%
%
\subsection{Learning Model Behavior through Model Exploration}
\label{subsec:method:exploration}
%
Like a human operator, Vito first learns how the model ``behaves'' by experimenting with it.
This is done through a ``self-guided sensitivity analysis''.
A batch of sample transitions is collected through model exploration episodes $\EpisodeSet^p = \{ \Episode_1^p, \Episode_2^p, \dots \}$.
An episode $\Episode_i^p$~is a sequence of \nEpisodeSteps~consecutive transitions generated from the model \CompModel~and the patient $p$ for whom the target measurements $\MeasurementVec^p$ are known.
An episode is initiated at time $t=0$ by generating random initial model parameters $\ParamVec_{t}$ within the physiologically plausible domain \ParamDomain.
From the outputs of a forward model run $\ModelStateVec_t = \CompModel(\ParamVec_{t})$, the misfits to the patient's corresponding measurements are computed, yielding the objectives vector $\ObjectiveVec_t = d(\ModelStateVec_t, \MeasurementVec^p)$.
Next, a random exploration policy $\Policy_\text{rand}$ that selects an action according to a discrete uniform probability distribution over the set of actions is employed.
The obtained $\Action_{t} \in \ActionSet$ is then applied to the current parameter vector, yielding modified parameter values $\ParamVec_{t+1} = \Action_t(\ParamVec_{t})$.
From the output of the forward model run $\ModelStateVec_{t+1} = \CompModel(\ParamVec_{t+1})$ the next objectives $\ObjectiveVec_{t+1}$ are computed.
The next action $\Action_{t+1}$ is then selected according to $\Policy_\text{rand}$, and this process is repeated $\nEpisodeSteps-1$ times.
Hence, each episode can be seen as a set of consecutive tuples:
\begin{equation}
  e = \{(\ParamVec_{t}, \ModelStateVec_{t}, \ObjectiveVec_{t}, \Action_{t}, \ParamVec_{t+1}, \ModelStateVec_{t+1}, \ObjectiveVec_{t+1}),\quad t=0, \dots, \nEpisodeSteps-1\} \enspace .
\end{equation}
In this work, $\nEpisodeSteps=100$ transitions are created in each episode as a trade-off between sufficient length of an episode to cover a real personalization scenario and sufficient exploration of the parameter space.

The model is explored with many different training patients and the resulting episodes are combined into one large training episode set $\EpisodeSet = \bigcup_p{\EpisodeSet^p}$.
The underlying hypothesis (verified in experiments) is that the combined \EpisodeSet~allows to cancel out peculiarities of individual patients, i.e.~to abstract from patient-specific to model-specific knowledge.
%
%
%
%
\subsection{From Computed Objectives to Representative MDP State}
\label{subsec:method:state_quantization}
\begin{figure}[t]
  \centering
  \includegraphics[width=.9\columnwidth]{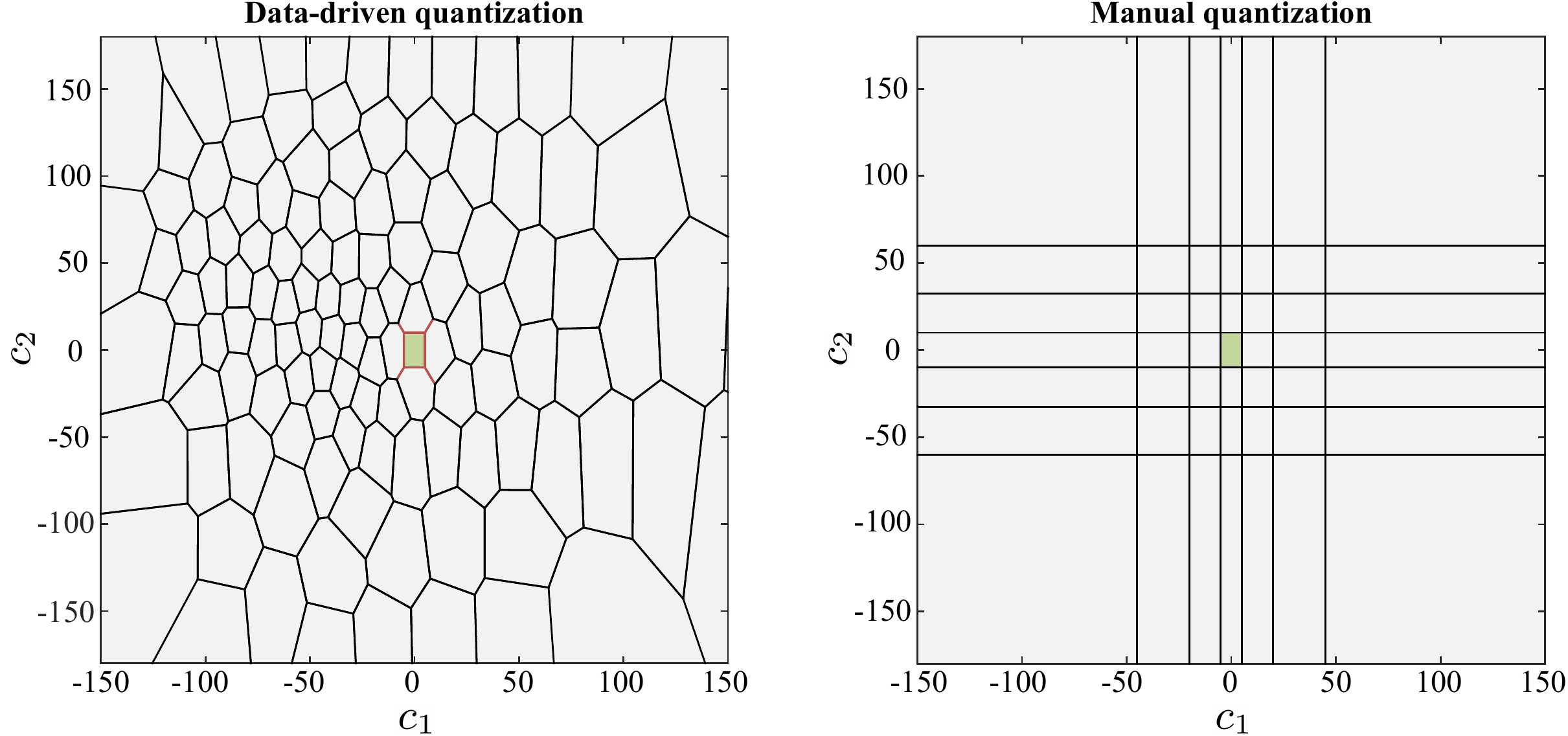}
  \caption{State-space quantization. \textbf{Left}: Example data-driven quantization of a two-dimensional state-space into $\nStates=120$ representative states.
  The states are distributed according to the observed objective vectors \ObjectiveVec~in one of the experiments in \autoref{sec:experiments:persoEP}. The objectives were QRS duration [ms] ($\Objective_1$) and electrical axis [deg] ($\Objective_2$).
  The center rectangle (green region) denotes the success state \StateOpt~\modified{where all objectives are met ($\forall i: |\Objective_i| < \Convergence_i$); see text for details}. 
  \textbf{Right}: Manual quantization as used in \cite{neumann2015vito}.}
  \label{fig:method:state_quantization}
\end{figure}

%
As mentioned above, the continuous space of objective vectors is quantized into a finite set of representative MDP states \StateSet.
A data-driven approach is proposed.
First, all objective vectors observed during training are clustered according to their distance to each other.
\modified{Because the ranges of possible values for the individual objectives can vary significantly depending on the selected measurements (due to different types of measurements, different units, etc.), the objectives should be normalized during clustering to avoid bias towards objectives with relatively large typical values.
In this work the distance measure performs implicit normalization to account for these differences: the distance between two objective vectors ($\ObjectiveVec_1, \ObjectiveVec_2$) is defined relative to} the inverse of the convergence criteria \ConvergenceVec:
\begin{equation}
  \|\ObjectiveVec_1 - \ObjectiveVec_2\|_{\ConvergenceVec} = 
  \sqrt{\left( \ObjectiveVec_1 - \ObjectiveVec_2 \right)^\top \diag(\ConvergenceVec)^{-1} \left( \ObjectiveVec_1 - \ObjectiveVec_2 \right)} \enspace ,
  \label{eq:method:distance_measure}
\end{equation}
where $\diag(\ConvergenceVec)^{-1}$ denotes a diagonal matrix with $(\frac{1}{\Convergence_1}, \frac{1}{\Convergence_2}, \dots)$ along its diagonal. 
The centroid of a cluster is the centroid of a representative state.
In addition, a special ``success state'' $\StateOpt$ representing personalization success is created, which covers the region in state-space where all objectives are met: $\forall i: |\Objective_i| < \Convergence_i$.
The full algorithm is described in \autoref{sec:appendix:quantization}.
Finally, an operator $\StateMapping: \mathbb{R}^\nObjectives \rightarrow \StateSet$ that maps continuous objective vectors \ObjectiveVec~to representative MDP states is introduced:
\begin{equation}
  \StateMapping(\ObjectiveVec) =
  \argmin_{\State \in \StateSet}
  \|\ObjectiveVec - \CentroidVec_\State\|_{\ConvergenceVec}
  \label{eq:method:mapping_phi}
\end{equation}
where $\CentroidVec_\State$ denotes the centroid corresponding to state \State.
For an example state-space quantization see \autoref{fig:method:state_quantization}.
%
%
%
\subsection{Transition Function as Probabilistic Model Representation}
\label{subsec:method:transitions}
%
In this work, the stochastic MDP transition function \Transitions~encodes the agent's knowledge about the computational model \CompModel. 
It is learned from the training data $\EpisodeSet$.
First, the individual samples $(\ParamVec_{t}, \ModelStateVec_{t}, \ObjectiveVec_{t}, \Action_{t}, \ParamVec_{t+1}, \ModelStateVec_{t+1}, \ObjectiveVec_{t+1})$ are converted to state-action-state transition tuples $\hat{\EpisodeSet} = \{(\State, \Action, \State')\}$, where $\State = \StateMapping(\ObjectiveVec_{t})$, $\Action = \Action_t$ and $\State' = \StateMapping(\ObjectiveVec_{t+1})$.
Then, \Transitions~is approximated from statistics over the observed transition samples:
\begin{equation}
  \Transitions(\State,\Action,\State') =
  \frac{
    \left\vert\{(\State,\Action,\State') \in \hat{\EpisodeSet}\}\right\vert
  }{
    \sum_{\State'' \in \StateSet}\left\vert\{(\State,\Action,\State'') \in \hat{\EpisodeSet}\}\right\vert
   }
  \enspace ,
  \label{eq:method:transition_probs}
\end{equation}
\modified{where $\left\vert\{\cdot\}\right\vert$ denotes the cardinality of the set $\{\cdot\}$.}
If \nStates~and \nActions~are large compared to the total number of samples it may occur that some state-action combinations are not observed: $|\{(\State, \Action, \cdot) \in \hat{\EpisodeSet}\}| = 0$.
In that case uniformity is assumed: $\forall \State'' \in \StateSet : \Transitions(\State,\Action,\State'') = 1/\nStates$.

\MDP~is now fully defined.
Value iteration (\autoref{subsec:method:mdp}) is applied and the stochastic policy $\PolicyStochasticOpt_\epsilon$ is computed, which completes the off-line phase.
%
%
%
\subsection{On-line Model Personalization}
\label{subsec:method:execution}
\begin{figure}[t]
  \centering
  \includegraphics[width=.9\columnwidth]{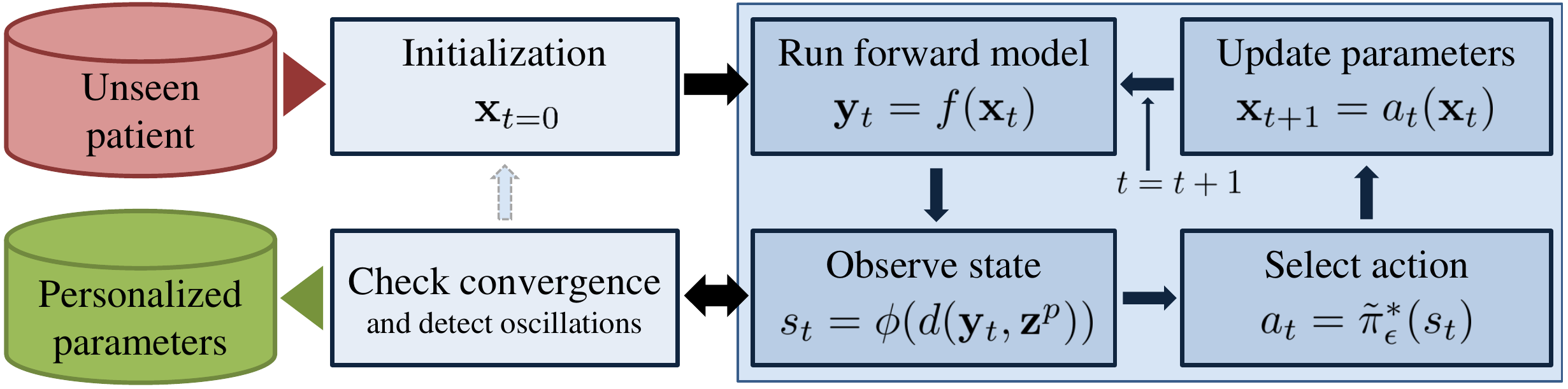}
  \caption{Vito's probabilistic on-line personalization phase. See text for details.}
  \label{fig:method:personalization}
\end{figure}

%
%
On-line personalization, as illustrated in \autoref{fig:method:personalization}, can be seen as a two-step procedure.
First, Vito initializes the personalization of unseen patients from training data.
Second, Vito relies on the computed policy $\PolicyStochasticOpt_\epsilon$~to guide the personalization process.
%
\subsubsection{Data-driven Initialization}
\label{subsubsec:method:initialization}
%
Good initialization can be decisive for a successful personalization.
Vito's strategy is to search for forward model runs in the training database \EpisodeSet~for which the model state $\CompModel(\ParamVec) = \ModelStateVec \approx \MeasurementVec^p$ is similar to the patient's measurements.
To this end, Vito examines all parameters $\Xi = \{\ParamVec \in \EpisodeSet \mid \CompModel(\ParamVec) \approx \MeasurementVec^p\}$ that yielded model states similar to the patient's measurements.
Due to ambiguities induced by the different training patients, data noise and model assumptions, $\Xi$ could contain significantly dissimilar parameters.
Hence, picking a single $\ParamVec \in \Xi$ might not yield the best initialization.
Analyzing $\Xi$ probabilistically instead helps Vito to find likely initialization candidates.
The details of the initialization procedure are described in \autoref{sec:appendix:initialization}.
Given the patient's measurements $\MeasurementVec^p$, the procedure outputs a list of initialization candidates \modified{$\mathcal{X}_0 = (\ParamVec_0', \ParamVec_0'', \dots)$. The list is sorted by likelihood with the first element, $\ParamVec_0'$, being the most likely one}.
%
%
%
\subsubsection{Probabilistic Personalization}
%
\modified{The first personalization step initializes the model parameter vector $\ParamVec_0$ with the most likely among all initialization candidates, $\ParamVec_0 \in \mathcal{X}_0$ (see previous section for details).
Then, }as illustrated in \autoref{fig:method:personalization}, Vito computes the forward model $\ModelStateVec_0 = \CompModel(\ParamVec_0)$ and the misfit between the model output and the patient's measurements $\ObjectiveVec_0 = d(\ModelStateVec_0, \MeasurementVec^p)$ to derive the first state $\State_0 = \StateMapping(\ObjectiveVec_0)$.
Given $\State_0$, Vito decides from its policy the first action to take $\Action_0 = \PolicyStochasticOpt_\epsilon(\State_0)$,
and walks through state-action-state sequences to personalize the computational model \CompModel~by iteratively updating the model parameters through MDP actions.
Bad initialization could lead to oscillations between states as observed in previous RL works~\citep{kveton2012kernel,neumann2015vito}.
Therefore, upon detection of an oscillation, which is done by monitoring the parameter traces to detect recurring sets of parameter values, the personalization is re-initialized at the second-most-likely $\ParamVec_0 \in \mathcal{X}_0$, etc.
If all $|\mathcal{X}_0|$ initialization candidates have been tested, a potential re-initialization defaults to fully random within the physiologically plausible parameter domain \ParamDomain.
The process terminates once Vito reaches state \StateOpt~(success), or when a pre-defined maximum number of iterations is reached (failure).
%
%
%
\section{Experiments}
\label{sec:experiments}
%
Vito was applied to a synthetic parameter estimation problem and to two challenging problems involving real clinical data: personalization of a cardiac electrophysiology (EP), and a whole-body-circulation (WBC) model.
All experiments were conducted using leave-one-out cross-validation.
\modified{The numbers of datasets and transition samples used for the different experiments are denoted $\nDatasets$ and \nTrainingSamples, respectively.}
%
%
\subsection{Synthetic Experiment: the Rosenbrock Function}
\label{sec:experiments:synthetic}
%
First, Vito was employed in a synthetic scenario, where the ground-truth model parameters were known.
The goals were to test the ability of Vito to optimize cost functions generically, and to directly evaluate the performance in the parameter space.
\begin{figure}[t]
  \centering
  \includegraphics[width=\columnwidth]{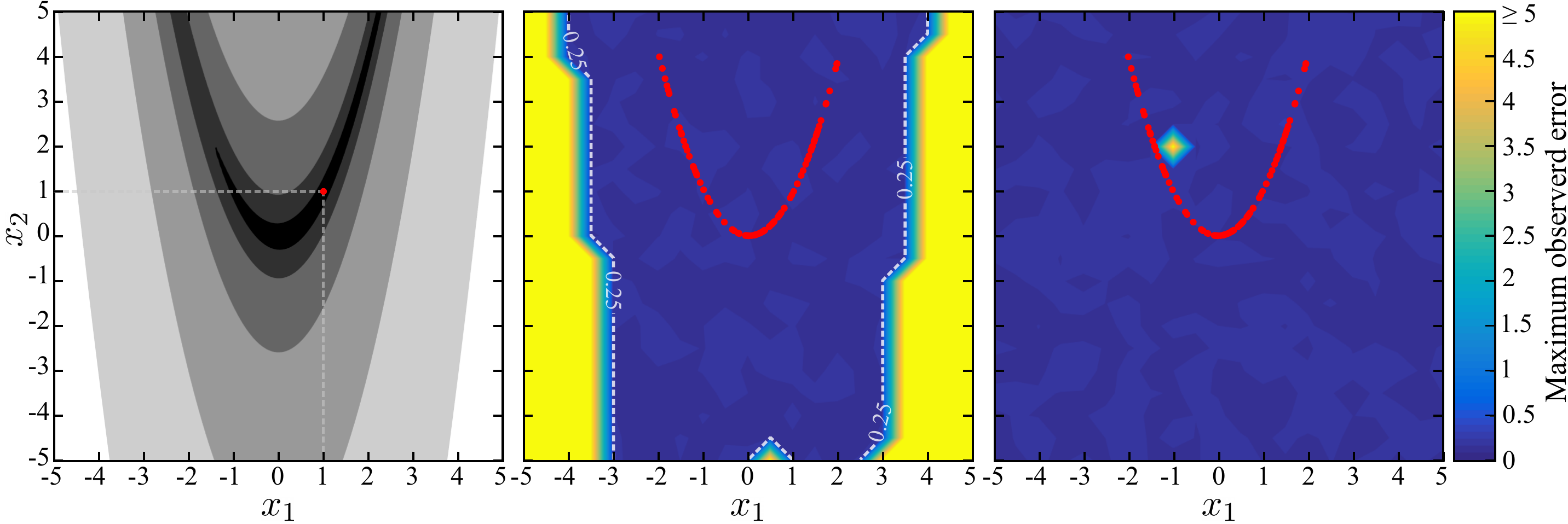}
  \caption{Synthetic experiment. \textbf{Left}: Contour plot of the Rosenbrock function $\CompModel^{\alpha=1}$ with global minimum at $\ParamVec = (1,1)^\top$ (red dot).
  The color scale is logarithmic for visualization purposes: the darker, the lower the function value.
  \textbf{Mid}: Maximum $L_2$-error in parameter space after personalization over all functions for varying initial parameter values. See text for details. Yellow represents errors $\geq 5$ (maximum observed error $\approx 110$).
  \textbf{Right}: Same as mid panel, except the extended action set was used.
  The red dots are the 100 ground-truth parameters $\ParamVec = (\alpha, \alpha^2)^\top$ generated for random $\alpha$.}
  \label{fig:experiments:rosen}
\end{figure}

\subsubsection{Forward Model Description} 
The Rosenbrock function \citep{rosenbrock1960automatic}, see \autoref{fig:experiments:rosen}, left panel, is a non-convex function that is often used to benchmark optimization algorithms.
It was treated as the forward model in this experiment:
\begin{equation}
  \CompModel^\alpha(\Param_1, \Param_2) = (\alpha - \Param_1)^2 + 100\cdot(\Param_2 - \Param_1^2)^2 \enspace ,
  \label{eq:experiments:rosenbrock}
\end{equation}
where $\ParamVec = (\Param_1, \Param_2)^\top$ were the model parameters to estimate for any $\alpha$, and $\CompModel^\alpha: \ParamDomain \rightarrow \mathbb{R}$.
As described in \autoref{subsubsec:method:reformulation}, each of Vito's actions $\Action \in \ActionSet$ in- or decrements a parameter value by multiples ($1\times,10\times,100\times$) of parameter-specific reference values.
The reference values were set to $\ReferenceValueVec = (0.01, 0.01)^\top$, determined as $0.1\%$ of the defined admissible parameter space per dimension, $\ParamDomain = [-5;5]^2$.
The parameter $\alpha \in \mathbb{R}$ defines a family of functions $\{\CompModel^\alpha\}$. 
The goal was to find generically $\argmin_{\Param_1, \Param_2} \CompModel^\alpha(\Param_1, \Param_2)$.

The Rosenbrock function has a unique global minimum at $\ParamVec=(\alpha,\alpha^2)^\top$, where both terms $T_1 = (\alpha-\Param_1)$ and $T_2 = (\Param_2-\Param_1^2)$ evaluate to 0. 
The personalization objectives were therefore defined as $\ObjectiveVec = (|T_1 - 0|, |T_2 - 0|)^\top$, with the measured data $\MeasurementVec = (0, 0)^\top$ were zero for both objectives and the computed data $\ModelStateVec = (T_1, T_2)^\top$.
The convergence criteria were set empirically to $\ConvergenceVec = (0.05, 0.05)^\top$. 

\subsubsection{Evaluation}
Vito was evaluated on $\nDatasets = 100$ functions $\CompModel^\alpha$ with randomly generated $\alpha \in [-2,2]$.
In the off-line phase, for each function, $\nTrainingSamples = 10\cdot \nEpisodeSteps = 1000$ samples, i.e.~ten training episodes, each consisting in $\nEpisodeSteps=100$ transitions (\autoref{subsec:method:exploration}), were generated to learn the policy.
The number of representative states was set to $\nStates = 100$.
To focus on Vito's on-line personalization capabilities, both the data-driven initialization and the re-initialization on oscillation (\autoref{subsec:method:execution}) were disabled.
In total, 441 experiments with different initializations (sampled on a $21 \times 21$ uniform grid spanned in $\ParamDomain$) were conducted.
For each experiment all 100 functions were personalized using leave-one-family-function-out cross validation, and the error value from the function exhibiting the maximum $L_2$-error (worst-case scenario) between ground-truth ($\alpha,\alpha^2$) and estimated parameters was plotted.
As one can see from the large blue region in \autoref{fig:experiments:rosen}, mid panel, for the majority of initial parameter values Vito always converged to the solution (maximum $L_2$-error $<0.25$; the maximum achievable accuracy depended on the specified convergence criteria \ConvergenceVec~and on the reference values \ReferenceValueVec, which ``discretized'' the parameter space).
However, especially for initializations far from the ground-truth (near border regions of $\ParamDomain$), Vito was unable to personalize some functions properly, which was likely due to the high similarity of the Rosenbrock function shape in these regions.

To investigate this issue, the experiment was repeated after additional larger parameter steps were added to the set of available actions: $\ActionSet' = \ActionSet \cup \{\pm 500 \ReferenceValue_1; \pm 500 \ReferenceValue_2\}$.
As shown in \autoref{fig:experiments:rosen}, right panel, Vito could now personalize successfully starting from any point in $\ParamDomain$.
The single spot with larger maximum error (bright spot at approximately $\ParamVec = (-1,2)^\top$) can be explained by Vito's stochastic behavior: Vito may have become \emph{unlucky} if it selected many unfavorable actions in sequence due to the randomness introduced by the stochastic policy.
Enabling re-initialization on oscillation solved this issue entirely.
In conclusion, this experiment showed that Vito can learn how to minimize a cost function generically. 
%
%
\subsection{Personalization of Cardiac Electrophysiology Model}
\label{sec:experiments:persoEP}
%
Vito was then tested in a scenario involving a complex model of cardiac electrophysiology coupled with 12-lead ECG.
Personalization was performed for real patients from actual clinical data.
A total of $\nDatasets = 83$ patients were available for experimentation.
For each patient, the end-diastolic bi-ventricular anatomy was segmented from short-axis cine magnetic resonance imaging (MRI) stacks as described in \cite{zheng2008four}.
A tetrahedral anatomical model including myofibers was estimated and a torso atlas affinely registered to the patient based on MRI scout images.
See \cite{zettinig2014data} for more details.
\subsubsection{Forward Model Description} 
The depolarization time at each node of the tetrahedral anatomical model was computed using a shortest-path graph-based algorithm, similar to the one proposed in~\cite{wallman2012comparative}.
Tissue anisotropy was modeled by modifying the edge costs to take into account fiber orientation.
A time-varying voltage map was then derived according to the depolarization time: at a given time $t$, mesh nodes whose depolarization time was higher than $t$ were assigned a trans-membrane potential of $-70$\,mV, $30$\,mV otherwise.
The time-varying potentials were then propagated to a torso model where 12-lead ECG acquisition was simulated, and QRS duration (QRSd) and electrical axis (EA) were derived \citep{zettinig2014data}.
The model was controlled by the conduction velocities (in m/s) of myocardial tissue and left and right Purkinje network: $\ParamVec = (v_\text{Myo}, v_\text{LV}, v_\text{RV})^\top$.
The latter two domains were modeled as fast endocardial conducting tissue.
The admissible parameter space $\ParamDomain$ was set to $[200;1000]$ for $v_\text{Myo}$ and $[500;5000]$ for both $v_\text{LV}$ and $v_\text{RV}$.
Reference increment values to build the action set $\ActionSet$ were set to $\ReferenceValueVec = (5, 5, 5)^\top$\,m/s for the three model parameters.
The goal of EP personalization was to estimate $\ParamVec$ from the measured QRSd and EA.
Accounting for uncertainty in the measurements and errors in the model, a patient was considered personalized if QRSd and EA misfits were below $\ConvergenceVec = (5\,\text{ms}, 10^\circ)^\top$, respectively.

\subsubsection{Number of Representative States}
\label{sec:exp:ep:num_rep_states}
\begin{figure*}[t]
  \centering
	\includegraphics[width=.9\columnwidth]{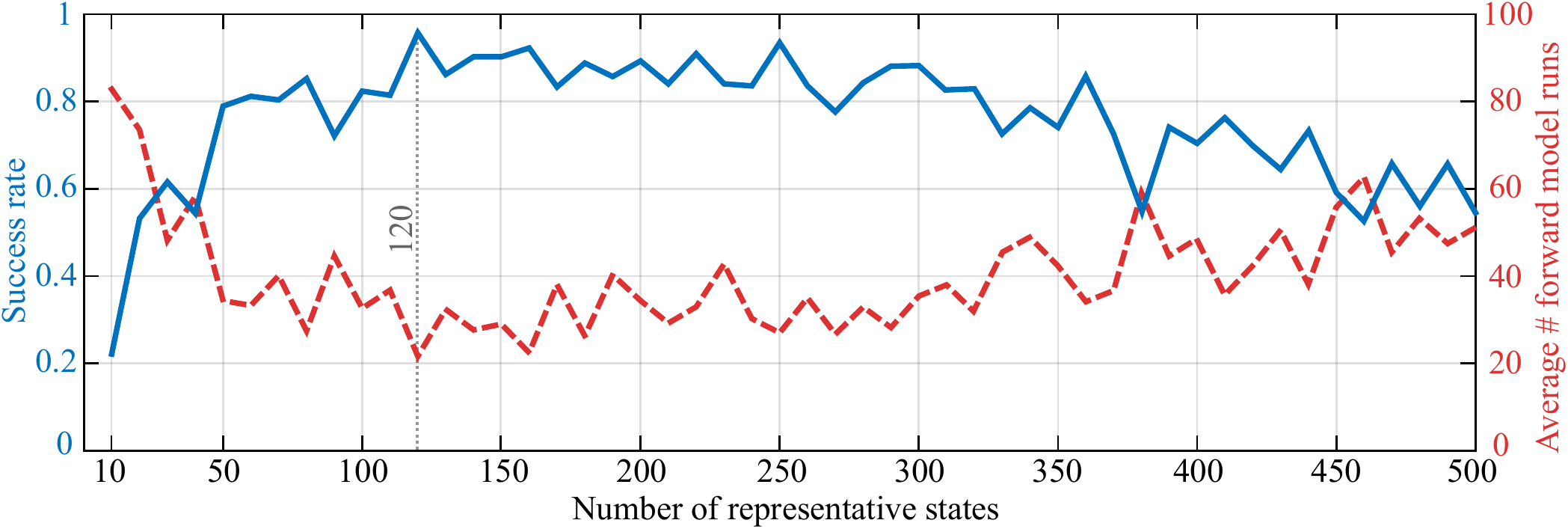}
  \caption{Hyper-parameter scouting. Vito's performance for varying number of representative states \nStates~on eight scouting datasets.
  \modified{The solid and dashed curves represent success rate and average number of forward runs until convergence, respectively, aggregated over five personalization runs with varying training data}.}
  \label{fig:method:ep_scouting}
\end{figure*}

In contrast to \cite{neumann2015vito}, where state-space quantization required manual tuning of various threshold values, the proposed approach relies on a single hyper-parameter only: \nStates, the number of representative states (\autoref{subsec:method:state_quantization}).
To specify \nStates, eight patients were selected for scouting.
Exhaustive search was performed for $\nStates \in \{10, 20, \dots, 490, 500 \}$ representative states.
The goodness of a given configuration was evaluated based on the success rate (relative number of successfully personalized cases according to convergence criteria $\ConvergenceVec$) over five independent, consecutive, leave-one-patient-out cross-validated personalization runs of the eight patients.
Furthermore, the average number of required forward model runs was considered.
To this end, 100 training episodes ($100 \cdot \nEpisodeSteps = 10^4$ transition samples) per patient were generated for each personalization run as described in \autoref{subsec:method:exploration}.
As one can see from \autoref{fig:method:ep_scouting}, good performance was achieved from 50 to 300 representative states.
The large range of well performing \nStates~indicates a certain level of robustness with respect to that hyper-parameter.
A slight performance peak at 120 representative states was observed.
Therefore, $\nStates=120$ was selected for further experimentation as compromise between maintaining a low number of states and sufficient state granularity.
An example quantization with $\nStates=120$ is visualized in \autoref{fig:method:state_quantization}.
The eight scouting datasets were discarded for the following experiments to avoid bias in the analysis.

\subsubsection{Reference Methods}
\label{subsubsec:exp:ep_reference_methods}
Vito's results were compared to two standard personalization methods based on BOBYQA \citep{powell2009bobyqa}, a widely-used gradient-free optimizer known for its robust performance and fast convergence.
The first approach, ``BOBYQA simple'', mimicked the most basic estimation setup, where only the minimum level of model and problem knowledge were assumed.
The objective function was the sum of absolute QRSd and EA errors: $\sum_{i=1}^\nObjectives |\Objective_i|$.
It was minimized in a single optimizer run where all three parameters in $\ParamVec$ were tuned simultaneously.
The algorithm terminated once all convergence criteria $\ConvergenceVec$ were satisfied (success) or if the number of forward model evaluations exceeded 100 (failure).
The second approach, ``BOBYQA cascade'', implemented an advanced estimator with strong focus on robustness, which computed the optimum parameters in a multi-step iterative fashion.
It is based on \cite{seegerer2015estimation}, where tedious manual algorithm and cost function tuning was performed on a subset of the data used in this manuscript.
In a first step, the myocardial conduction velocity was tuned to yield good match between computed and measured QRS duration.
Second, left and right endocardial conduction velocities were optimized to minimize electrical axis error.
Both steps were repeated until all parameter estimates were stable.

To remove bias towards the choice of initial parameter values, for each of the two methods all datasets were personalized 100 times with different random initializations within the range of physiologically plausible values $\ParamDomain$.
The differences in performance were striking: only by changing the initialization, the number of successfully personalized cases varied from 13 to 37 for BOBYQA simple, and from 31 to 51 for BOBYQA cascade (variability of more than 25\% of the total number of patients).
These results highlight the non-convexity of the cost function to minimize.

\subsubsection{Full Personalization Performance}
\begin{figure}
	\centering
	\includegraphics[width=\columnwidth]{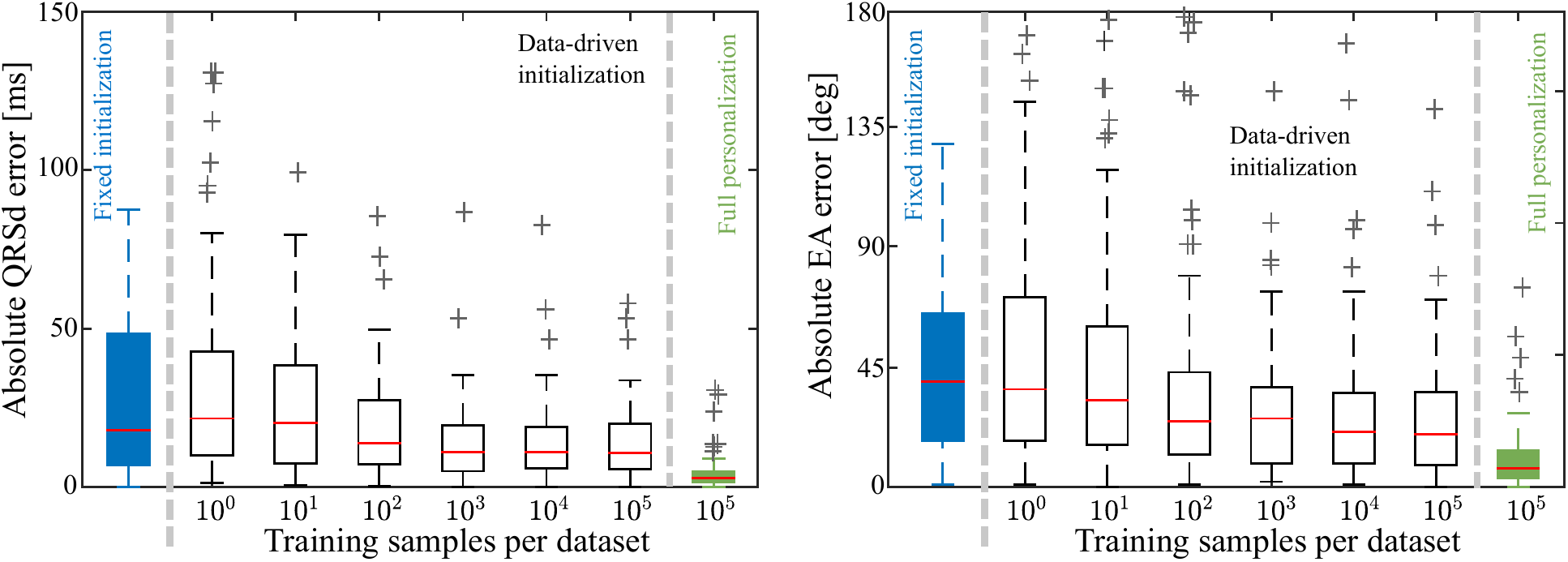}
	\caption{Absolute errors over all patients after initialization with fixed parameter values (blue), after data-driven initialization for increasing amount of training data (white), and after full personalization with Vito (green).
  Data-driven initialization yielded significantly reduced errors if sufficient training data were available ($>10^2$) compared to initialization with fixed values.
  Full personalization further reduced the errors by a significant margin.
  The red bar and the box edges indicate the median absolute error, and the 25 and 75 percentiles, respectively.
  \textbf{Left}: QRS duration errors.
  \textbf{Right}: Electrical axis errors.}
	\label{fig:experiments:error_after_initialization}
\end{figure}

First, Vito's overall performance was evaluated.
The full personalization pipeline consisting in off-line learning, initialization, and on-line personalization was run on all patients with leave-one-patient-out cross-validation using 1000 training episodes ($\nTrainingSamples = 1000\cdot \nEpisodeSteps = 10^5$ transition samples) per patient.
The maximum number of iterations was set to 100.
The green box plots in the two panels of \autoref{fig:experiments:error_after_initialization} summarize the results.
The mean absolute errors were $4.1 \pm 5.6$\,ms and $12.4 \pm 13.3^\circ$ in terms of QRSd and EA, respectively, a significant improvement over the residual error after initialization.
In comparison to the reference methods, the best BOBYQA simple run yielded absolute errors of $4.4 \pm 10.8$\,ms QRSd and $15.5 \pm 18.6^\circ$ EA on average, and the best BOBYQA cascade run $0.1 \pm 0.2$\,ms QRSd and $11.2 \pm 15.8^\circ$ EA, respectively.
Thus, in terms of EA error all three methods yielded comparable performance, and while BOBYQA simple and Vito performed similarly in terms of QRSd, BOBYQA cascade outperformed both in this regard.
However, considering success rates, i.e.~successfully personalized patients according to the defined convergence criteria ($\ConvergenceVec$) divided by total number of patients, both the performance of Vito (67\%) and BOBYQA cascade (68\%) were equivalent, while BOBYQA simple reached only 49\% or less.
In terms of run-time, i.e.~average number of forward model runs until convergence, Vito (31.8) almost reached the high efficiency of BOBYQA simple (best: 20.1 iterations) and clearly outperformed BOBYQA cascade (best: 86.6 iterations), which means Vito was $\approx 2.5\times$ faster.

\subsubsection{Residual Error after Initialization}
A major advantage over standard methods such as the two BOBYQA approaches is Vito's automated, data-driven initialization method (\autoref{subsubsec:method:initialization}), which eliminates the need for user-provided initial parameter values.
To evaluate the utility of this step, personalization using Vito was stopped directly after initialization (the most likely $\ParamVec_0$ was used) and the errors in terms of QRSd and EA resulting from a forward model run $\CompModel$ with the computed initial parameter values were quantified.
This experiment was repeated for increasing number of transition samples per dataset: $\nTrainingSamples = 10^0 \dots 10^5$, and the results were compared to the error after initialization when fixed initial values were used (the initialization of the best performing BOBYQA experiment was used).
As one can see from \autoref{fig:experiments:error_after_initialization}, with increasing amount of training data both errors decreased notably.
As few as $10^2$ transitions per dataset already provided more accurate initialization than the best tested fixed initial values.
Thus, not only does this procedure simplify the setup of Vito for new problems (no user-defined initialization needed), this experiment showed that it can reduce initial errors by a large margin, even when only few training transitions were available.
It should be noted that Vito further improves the model fit in its normal operating mode (continue personalization after initialization), as shown in the previous experiment.

\subsubsection{Convergence Analysis}
\label{sec:experiments:persoEP_conv}
\begin{figure}
  \centering
  \includegraphics[width=\columnwidth]{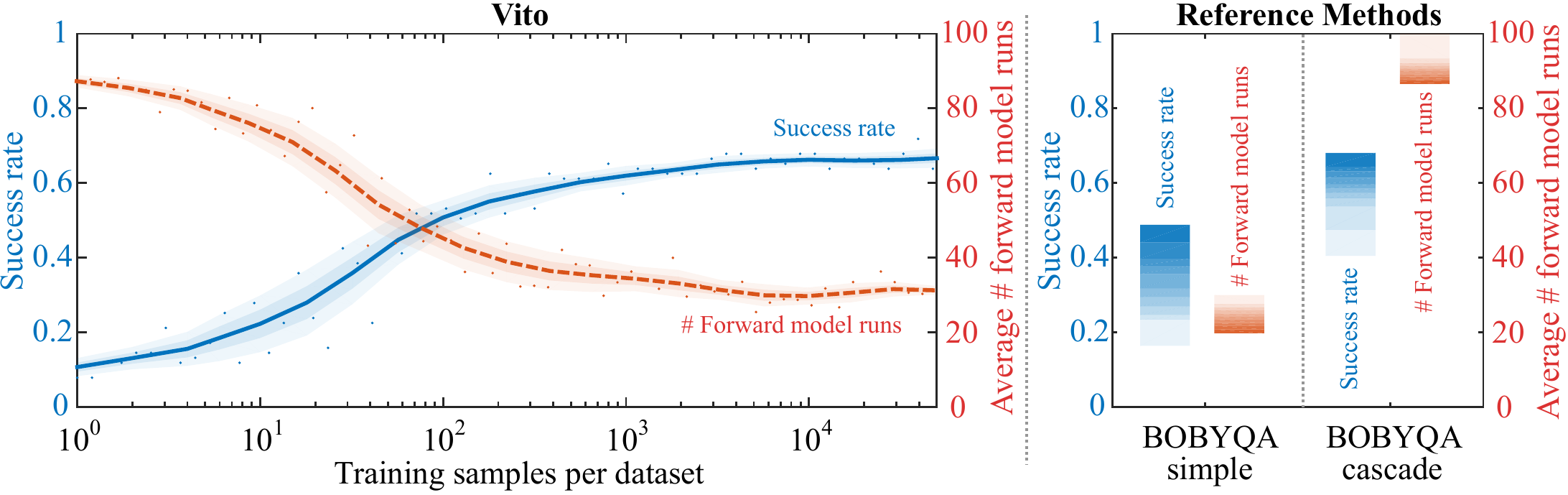}
  \caption{EP personalization results. Personalization success rate in blue and average number of iterations in red.
  \textbf{Left}: Vito's performance for increasing number of training transition samples per dataset. Each dot represents results from one experiment (cross-validated personalization of all 75 datasets), solid/dashed line is low-pass filtered mean, shaded areas represent $0.5\times$ and $1\times$ standard deviation.
  \textbf{Right}: Performance of both reference methods. Each shade represents $10\%$ of the results, sorted by performance.}
  \label{fig:experiments:performance_vs_samples}
\end{figure}

An important question in any RL application relates to the amount of training needed until convergence of the artificial agent's behavior.
For Vito in particular, this translates to the amount of transition samples required to accurately estimate the MDP transition function \Transitions~to compute a solid policy on the one hand, and to have enough training data for reliable parameter initialization on the other hand.
To this end, Vito's overall performance (off-line learning, initialization, personalization) was evaluated for varying number of training transition samples per dataset.
As one can see from the results in \autoref{fig:experiments:performance_vs_samples}, with increasing amount of training data the performance increased, suggesting that the learning process was working properly.
Even with relatively limited training data of only $\nTrainingSamples = 10^2$ samples per patient, Vito outperformed the best version of BOBYQA simple ($49\%$ success rate).
Starting from $\nTrainingSamples \approx 3000$, a plateau at $\approx$66\% success rate was reached, which remained approximately constant until the maximum tested number of samples.
This was almost on par with the top BOBYQA cascade performance ($68\%$ success rate).
Also the run-time performance increased with more training data.
For instance, Vito's average number of iterations was 36.2 at $10^3$ samples, 31.5 at $10^4$ samples, or 31.8 at $10^5$ samples.

These results suggested that not only Vito can achieve similar performance as an advanced, manually engineered method, but also the number of required training samples was not excessive.
In fact, a rather limited and thus well manageable amount of data, which can be computed in a reasonable time-frame, sufficed.
%
%
%
\subsection{Personalization of Whole-Body Circulation Model}
\label{sec:experiments:persoWBC}
%
Next, Vito was asked to personalize a lumped-parameter whole-body circulation (WBC) model from pressure catheterization and volume data.
A subset of $\nDatasets = 56$ patients from the EP experiments were used for experimentation.
The discrepancy was due to missing catheterization data for some patients, which was required for WBC personalization only.
For each patient, the bi-ventricular anatomy was segmented and tracked from short-axis cine MRI stacks throughout one full heart cycle using shape-constraints, learned motion models and diffeomorphic registration \citep{wang2013learning}.
From the time-varying endocardial meshes, ventricular volume curves were derived.
Manual editing was performed whenever necessary.

\subsubsection{Forward Model Description}
The WBC model to personalize was based on \cite{itu2014parestimation}.
It contained a heart model (left ventricle (LV) and atrium, right ventricle and atrium, valves), the systemic circulation (arteries, capillaries, veins) and the pulmonary circulation (arteries, capillaries, veins).
Time-varying elastance models were used for all four chambers of the heart.
The valves were modeled through a resistance and an inertance.
A three-element Windkessel model was used for the systemic and pulmonary arterial circulation, while a two-element Windkessel model was used for the systemic and pulmonary venous circulation.
We refer the reader to \cite{itu2014parestimation,neumann2015vito,westerhof1971artificial} for more details.
Personalization was performed with respect to the patient's heart rate as measured during catheterization.

\begin{figure}
  \centering
  \includegraphics[width=\columnwidth]{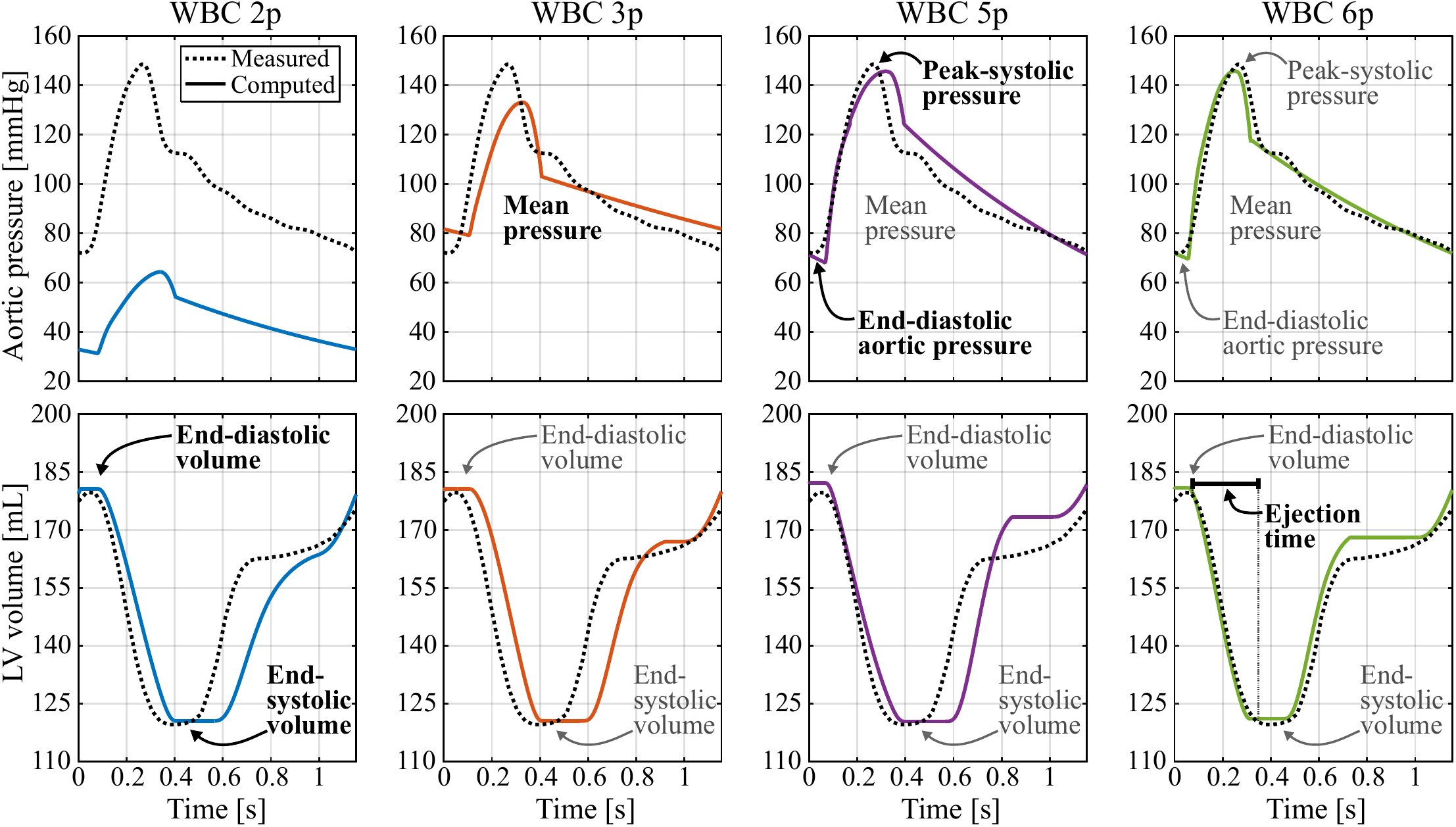}
  \caption{Goodness of fit in terms of time-varying LV volume and aortic pressure for Vito personalizing an example patient based on the different WBC setups.
  The added objectives per setup are highlighted in the respective column.
  With increasing number of parameters and objectives Vito manages to improve the fit between model and measurements.}
  \label{fig:experiments:wbc_eval_setups}
\end{figure}

The goal of this experiment was to compare Vito's personalization performance for the systemic part of the model in setups with increasing number of parameters to tune and objectives to match.
To this end, Vito was employed on setups with two to six parameters (2p, 3p, 5p, 6p): initial blood volume, LV maximum elastance, time until maximum elastance is reached, total aortic resistance and compliance, and LV dead volume.
The reference values $\ReferenceValueVec$ to define Vito's allowed actions $\ActionSet$ were set to $.5\%$ of the admissible parameter range $\ParamDomain$ for each individual parameter, see \autoref{tab:exp:wbc_params} for details.
\afterpage{
\begin{table}[t]
  \centering
  \begin{tabular}{|l|l|l||l|}
    \hline
    $\ParamVec$ & Default value & $\ParamDomain$ & \textit{Setups}\\\hline
    Initial volume & 400\,\small{mL} & $[200; 1000]$ \small{mL} & 6, 5, 3, 2 \\
    LV max.~elastance & 2.4\,\small{mmHg/mL} & $[0.2; 5]$ \small{mmHg/mL} & 6, 5, 3, 2 \\
    Aortic resistance & 1100\,\small{g$/$(cm$^4$\,s)} & $[500; 2500]$ \small{g$/$(cm$^4$\,s)} & 6, 5, 3 \\
    Aortic compliance & 1.4\,\small{$\cdot 10^9$\,cm$^4$\,s$^2/$g} & $[0.5; 6]$ \small{$\cdot 10^9$\,cm$^4$\,s$^2/$g} & 6, 5 \\
    Dead volume & 10\,\small{mL} & $[-50; 500]$ \small{mL} & 6, 5 \\
    Time to $E_\text{max}$ & 300\,\small{ms} & $[100; 600]$ \small{ms} & 6 \\
    \hline
  \end{tabular}
  \caption{WBC parameters $\ParamVec$, their default values and domain $\ParamDomain$. The last column denotes the experiment setups in which a parameter was personalized (e.g.~``5'': parameter was among the estimated parameters in 5p experiment). Default values were used in experiments where the respective parameters were not personalized.}
  \label{tab:exp:wbc_params}
\end{table}
\begin{table}[t]
  \centering
  \begin{tabular}{|l|l|l||l|}
    \hline
    $\ObjectiveVec$ & $\ConvergenceVec$ & Measured range & \textit{Setups} \\\hline
    End-diastolic LV volume & 20\,\small{mL} & $[129; 647]$ \small{mL}& 6, 5, 3, 2 \\
    End-systolic LV volume & 20\,\small{mL} & $[63; 529]$ \small{mL} & 6, 5, 3, 2 \\
    Mean aortic pressure & 10\,\small{mmHg} & $[68; 121]$ \small{mmHg} & 6, 5, 3 \\
    Peak-systolic aortic pressure & 10\,\small{mmHg} & $[83; 182]$ \small{mmHg} & 6, 5 \\
    End-diastolic aortic pressure & 10\,\small{mmHg} & $[48; 99]$ \small{mmHg} & 6, 5 \\
    Ejection time & 50\,\small{ms} & $[115; 514]$ \small{ms} & 6 \\
    \hline
  \end{tabular}
  \caption{WBC objectives $\ObjectiveVec$, their convergence criteria $\ConvergenceVec$ and range of measured values in the patient population used for experimentation.}
  \label{tab:exp:wbc_objectives}
\end{table}
}
The personalization objectives were MRI-derived end-diastolic and end-systolic LV volume, ejection time (time duration during which the aortic valve is open and blood is ejected), and peak-systolic, end-diastolic, and mean aortic blood pressures as measured during cardiac catheterization, see \autoref{fig:experiments:wbc_eval_setups}.
To account for measurement noise, personalization was considered successful if the misfits per objective were below acceptable threshold values $\ConvergenceVec$ as listed in \autoref{tab:exp:wbc_objectives}.

\subsubsection{Number of Representative States}
Along the same lines as \autoref{sec:exp:ep:num_rep_states}, the hyper-parameter for state-space quantization was tuned based on the eight scouting patients.
The larger the dimensionality of the state-space, the more representative states were needed to yield good performance.
In particular, for the different WBC setups, the numbers of representative states (\nStates) yielding the best scouting performance were 70, 150, 400 and 600 for the 2p, 3p, 5p and 6p setup, respectively.
The scouting datasets were discarded for the following experiments.

\subsubsection{Reference Method}
A gradient-free optimizer \citep{lagarias1998convergence} based on the simplex method was used to benchmark Vito.
The objective function was the sum of squared differences between computed and measured values, weighted by the inverse of the convergence criteria \modified{to counter the different ranges of objective values (e.g.~due to different types of measurements and different units): $\|\ObjectiveVec\|_{\ConvergenceVec}$ (\autoref{eq:method:distance_measure}).
Compared to non-normalized optimization, the algorithm converged up to 20\% faster and success rates increased by up to 8\% under otherwise identical conditions.}
Personalization was terminated once all convergence criteria were satisfied (success), or when the maximum number of iterations was reached (failure).
To account for the increasing complexity of optimization with increasing number of parameters \nParams, the maximum number of iterations was set to $50 \cdot \nParams$ for the different setups.

\begin{figure}
  \centering
  \includegraphics[width=\columnwidth]{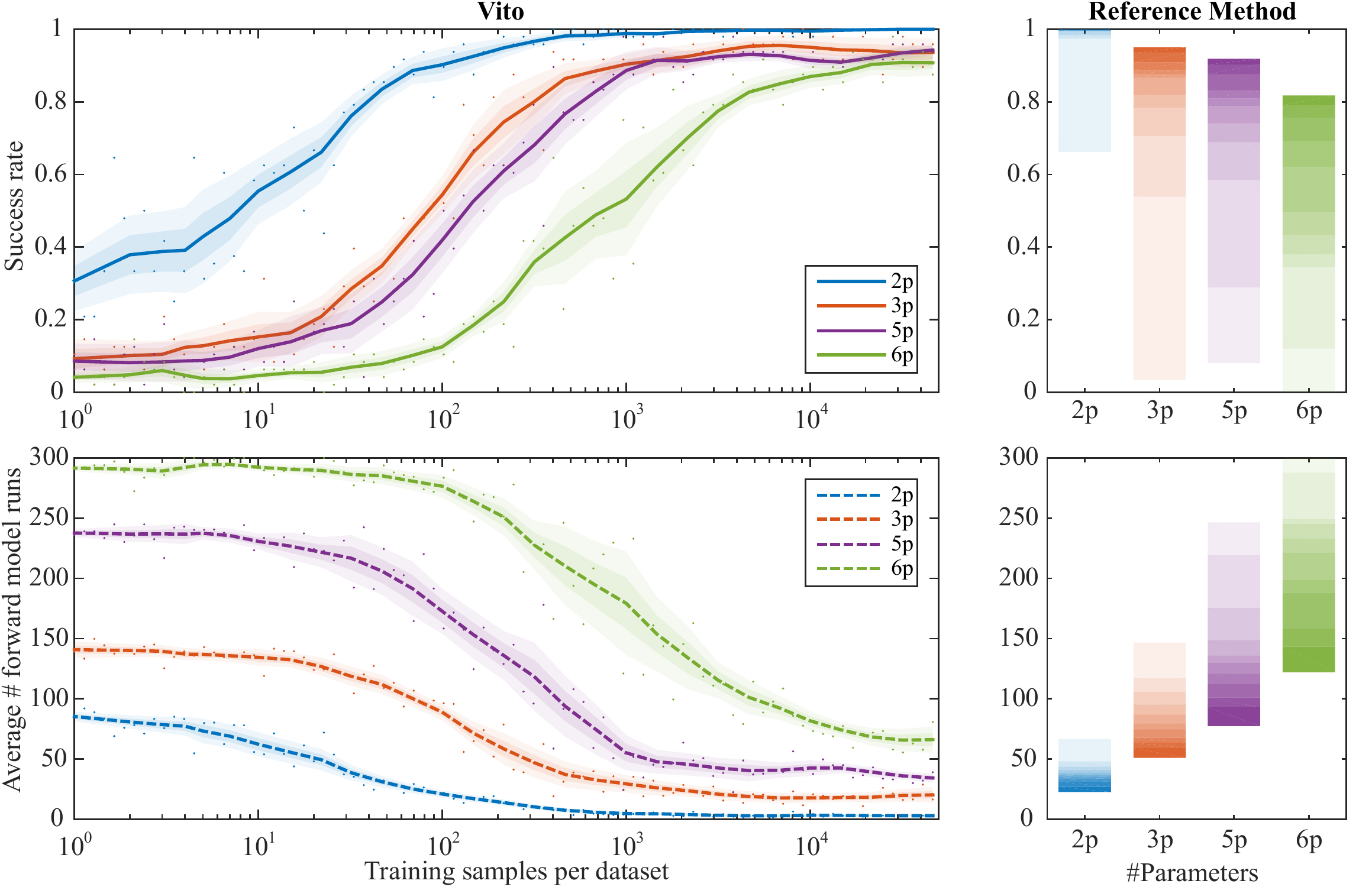}
  \caption{WBC model personalization results (top: success rate, bottom: average number of forward model runs until convergence) for various estimation setups (different colors), see text for details.
  \textbf{Left}: Vito's performance for increasing number of training transition samples per dataset. Each dot represents results from one experiment (cross-validated personalization of all 48 datasets), solid/dashed lines are low-pass filtered mean, shaded areas represent $0.5\times$ and $1\times$ standard deviation.
  \textbf{Right}: Performance of reference method. Each shade represents $10\%$ of the results, sorted by performance; darkest shade: best $10\%$.}
  \label{fig:experiments:wbc_plots}
\end{figure}

As one can see from \autoref{fig:experiments:wbc_plots}, right panels, with increasing number of parameters to be estimated, the performance in terms of success rate and number of forward model runs decreased slightly.
This is expected as the problem becomes harder.
To suppress bias originating from (potentially poor) initialization, the reference method was run 100 times per setup (as in EP experiments), each time with a different, randomly generated set of initial parameter values.
The individual performances varied significantly for all setups.

\subsubsection{Convergence Analysis}
\label{sec:experiments:persoWBC_conv}
For each WBC setup the full Vito personalization pipeline was evaluated for increasing training data ($\nTrainingSamples = 10^0 \dots 10^5$) using leave-one-patient-out cross-validation.
The same iteration limits as for the reference method were used.
The results are presented in \autoref{fig:experiments:wbc_plots}, left panels.
With increasing data, Vito's performance, both in terms of success rate and run-time (iterations until convergence), increased steadily until reaching a plateau.
As one would expect, the more complex the problem, i.e.~the more parameters and objectives involved in the personalization, the more training data was needed to reach the same level of performance.
For instance, Vito reached $80\%$ success rate with less than $\nTrainingSamples=50$ training samples per dataset in the 2p setup, whereas almost 90$\times$ as many samples were required to achieve the same performance in the 6p setup. 

Compared to the reference method, given enough training data, Vito reached equivalent or better success rates (e.g.~up to 11\%~higher success rate for 6p) while significantly outperforming the reference method in terms of run-time performance.
In the most basic setup (2p), if $\nTrainingSamples \geq 10^3$, Vito converged after $3.0$ iterations on average, while the best reference method run required $22.6$ iterations on average, i.e.~Vito was seven times faster.
For the more complex setups (3p, 5p, 6p), the speed-up was not as drastic.
Yet, in all cases Vito outperformed even the best run of the reference method by a factor of $1.8$ or larger. 
%
%
%
\section{Conclusion}
\label{sec:conclusion}
%
%
\subsection{Summary and Discussion}
%
In this manuscript, a novel personalization approach called Vito has been presented.
To our knowledge, it is the first time that \modified{biophysical model personalization} is addressed using artificial intelligence concepts.
Inspired by how humans approach the personalization problem, Vito first learns the characteristics of the computational model under consideration using a data-driven approach.
This knowledge is then utilized to learn how to personalize the model using reinforcement learning.
Vito is generic in the sense that it requires only minimal and intuitive user input (parameter ranges, authorized actions, number of representative states) to learn by itself how to personalize a model.

Vito was applied to a synthetic scenario and to two challenging personalization tasks in cardiac computational modeling.
The problem setups and hyper-parameter configurations are listed in \autoref{tab:concl:setups}.
In most setups the majority of hyper-parameters were identical and only few (\nStates) required manual tuning, suggesting good generalization properties of Vito.
\begin{table}[t]
  \centering
  \begin{tabular}{|l||c|c|c||r|c|r|}
    \hline
    \textbf{Application} & \nParams & \nObjectives & \nDatasets & \nStates & $\nActions/\nParams$ & $n_\text{plateau}$
    \\\hline
    Rosenbrock & 2 & 2 & 100 & 100 & 6 & \textit{n/a} \\
    Rosenbrock ext. & 2 & 2 & 100 & 100 & 8 & \textit{n/a} \\
    EP & 3 & 2 & 83 (75) & 120 & 6 & 3\,000 \\
    WBC 2p & 2 & 2 & 56 (48) & 70 & 6 & 450 \\
    WBC 3p & 3 & 3 & 56 (48) & 150 & 6 & 2\,000 \\
    WBC 5p & 5 & 5 & 56 (48) & 400 & 6 & 3\,500 \\
    WBC 6p & 6 & 6 & 56 (48) & 600 & 6 & 20\,000 \\
    \hline
  \end{tabular}
  \caption{Applications considered in this manuscript described in terms of the number of parameters (\nParams), objectives (\nObjectives) and datasets (\nDatasets) used for experimentation (in brackets: excluding scouting patients, if applicable); and Vito's hyper-parameters: the number of representative MDP states (\nStates) and the number of actions per parameter ($\nActions/\nParams$). The last column ($n_\text{plateau}$) denotes the approximate number of samples needed to reach the performance ``plateau'' (see convergence analyses in \autoref{sec:experiments:persoEP_conv} and \autoref{sec:experiments:persoWBC_conv}).}
  \label{tab:concl:setups}
\end{table}
Another key result was that Vito was up to 11\% more robust (higher success rates) compared to standard personalization methods.
Vito's ability to generalize the knowledge obtained from a set of training patients to personalize unseen patients was shown as all experiments reported in this manuscript were based on cross-validation.
Furthermore, Vito's robustness against training patients for whom we could not find a solution was tested.
In particular, for about 20\% of the patients, in none of the electrophysiology experiments in \autoref{sec:experiments:persoEP} any  personalization (neither Vito nor the reference methods) could produce a result that satisfied all convergence criteria.
Hence, for some patients no solution may exist under the given electrophysiology model configuration\footnote{Potential solution non-existence may be due to possibly invalid assumptions of the employed EP model for patients with complex pathologies.}.
Still, all patients were used to train Vito, and surprisingly Vito was able to achieve equivalent success rate as the manually engineered personalization approach for cardiac EP.

Generating training data could be considered Vito's computational bottleneck.
However, training is \textit{i)} performed off-line and one-time only, and \textit{ii)} it is independent for each training episode and each patient.
Therefore, \modified{large computing clusters} could be employed to perform rapid training by parallelizing this phase.
On-line personalization, on the contrary, is not parallelizable in its current form: the parameters for each forward model run depend on the outcome of the previous iteration.
Since the forward computations are the same for every ``standard'' personalization method (not including surrogate-based approaches), the number of forward model runs until convergence was used for benchmarking: Vito was up to seven times faster compared to the reference methods.
The on-line overhead introduced by Vito (convert data into an MDP state, then query policy) is negligible.

As such, Vito could become a unified framework for personalization of any computational physiological model, potentially eliminating the need for an expert operator with in-depth knowledge to design and engineer complex optimization procedures.
%
\subsection{Challenges and Outlook}
%
Important challenges still remain, such as the incorporation of continuous actions, the definition of states and their quantization.
In this work we propose a data-driven state-space quantization strategy.
Contrary to \cite{neumann2015vito}, where a threshold-based state-quantization involving several manually tuned threshold values (\autoref{fig:method:state_quantization}) was employed, the new method is based on a single hyper-parameter only: the number of representative states.
Although it simplifies the setup of Vito, this quantization strategy may still not be optimal, especially if only little training data is available.
Therefore, advanced approaches for continuous reinforcement learning with value function approximation \citep{sutton1998reinforcement,mnih2015human} could be integrated to fully circumvent quantization issues.

At the same time, such methods could improve Vito's scalability towards high-dimensional estimation tasks.
In this work we showed that Vito can be applied to typical problems emerging in cardiac modeling, which could be described as medium-scale problems with moderate number of parameters to personalize and objectives to match.
In unreported experiments involving $>$10 parameters, however, Vito could no longer reach satisfactory performance, which is likely due to the steeply increasing number of transition samples needed to sample the continuous state-space of increasing dimensionality sufficiently during training.
The trends in \autoref{sec:experiments:persoWBC} confirm the need for more data.
In the future, experience replay \citep{lin1993reinforcement,adam2012experience} or similar techniques could be employed to increase training data efficiency.
Furthermore, massively parallel approaches \citep{nair2015massively} are starting to emerge, opening up new avenues for large-scale reinforcement learning.

Although the employed reinforcement learning techniques guarantee convergence to an optimal policy, the computed personalization strategy may not be optimal for the model under consideration as the environment is only partially observable and the personalization problem ill-posed: there is no guarantee for solution existence or uniqueness.
Yet, we showed that Vito can solve personalization more robustly and more effectively than standard methods under the same conditions.
However, a theoretical analysis in terms of convergence guarantees and general stability of the method would be desirable, in particular with regards to the proposed re-initialization strategy.
As a first step towards this goal, in preliminary (unreported) experiments on the EP and the WBC model we observed that the number of patients which do not require re-initialization (due to oscillation) to converge to a successful personalization consistently increased with increasing training data.

\modified{The data-driven initialization proposed in this work simplifies Vito's setup by eliminating the need for user-provided initialization.
However, currently there is no guarantee that the first initialization candidate is the one that will yield the ``best'' personalization outcome.
Therefore, one could investigate the benefits of a fuzzy personalization scheme: many personalization processes could be run in parallel starting from the different initialization candidates.
Parameter uncertainty quantification techniques \citep{neumann2014robust} could then be applied to compute a probability density function over the space of model parameters.
Such approaches aim to gather complete information about the solution-space, which can be used to study solution uniqueness and other interesting properties.}

\modified{An important characteristic of any personalization algorithm is its stability against small variations of the measured data.
A preliminary experiment indicated good stability of Vito: the computed parameters from several personalization runs, each involving small random perturbations of the measurements, were consistent.
Yet in a small group of patients some parameter variability was observed, however, it was below the variability of the reference method under the same conditions. 
To what extent certain degrees of variability will impact other properties of the personalized model such as its predictive power will be subject of future research. 
We will also investigate strategies to improve Vito's stability further.
For instance, the granularity of the state-space could provide some flexibility to tune the stability: less representative states means a larger region in state space per state, thus small variations in the measured data might have less impact on personalization outcome.
However, this could in turn have undesirable effects on other properties of Vito such as success rate or convergence speed (see \autoref{sec:exp:ep:num_rep_states}).}

Beyond these challenges, Vito showed promising performance and versatility, making it a first step towards an automated, self-taught model personalization agent.
The next step will be to investigate the predictive power of the personalized models, for instance for predicting acute or long-term response in cardiac resynchronization therapy \citep{sermesant2009personalised,kayvanpour2015towards}.
%
%
%
%
\appendix
%
\section{Data-driven State-Space Quantization}
\label{sec:appendix:quantization}
%
This section describes the details of the proposed data-driven quantization approach to define the set of representative MDP states $\StateSet$ (see \autoref{subsec:method:state_quantization}).
It is based on clustering, in particular on the weighted \textit{k}-means algorithm described in \cite{arthur2007k}.
To this end, all objective vectors $\ObjectiveSet = \{\ObjectiveVec \in \EpisodeSet\}$ are extracted from the training data (\autoref{subsec:method:exploration}).
$\ObjectiveSet \subset \mathbb{R}^\nObjectives$ represents all observed ``continuous states''.
The goal is to convert $\ObjectiveSet$ into the finite set of representative MDP states $\StateSet$ while taking into account that Vito relies on a special ``success state'' \StateOpt~encoding personalization success.

The success state \StateOpt~does not depend on the data, but on the maximum acceptable misfit \ConvergenceVec.
In particular, since personalization success implies that all objectives are met, \StateOpt~should approximate a hyperrectangle centered at $\vec{0}$ and bounded at $\pm \ConvergenceVec$, i.e.~a small region in $\mathbb{R}^\nObjectives$ where $\forall i: |\Objective_i| < \Convergence_i$.
To enforce $\StateOpt$, the input to weighted \textit{k}-means is preprocessed as follows.

\begin{figure}[t]
  \centering
  \includegraphics[width=.9\columnwidth]{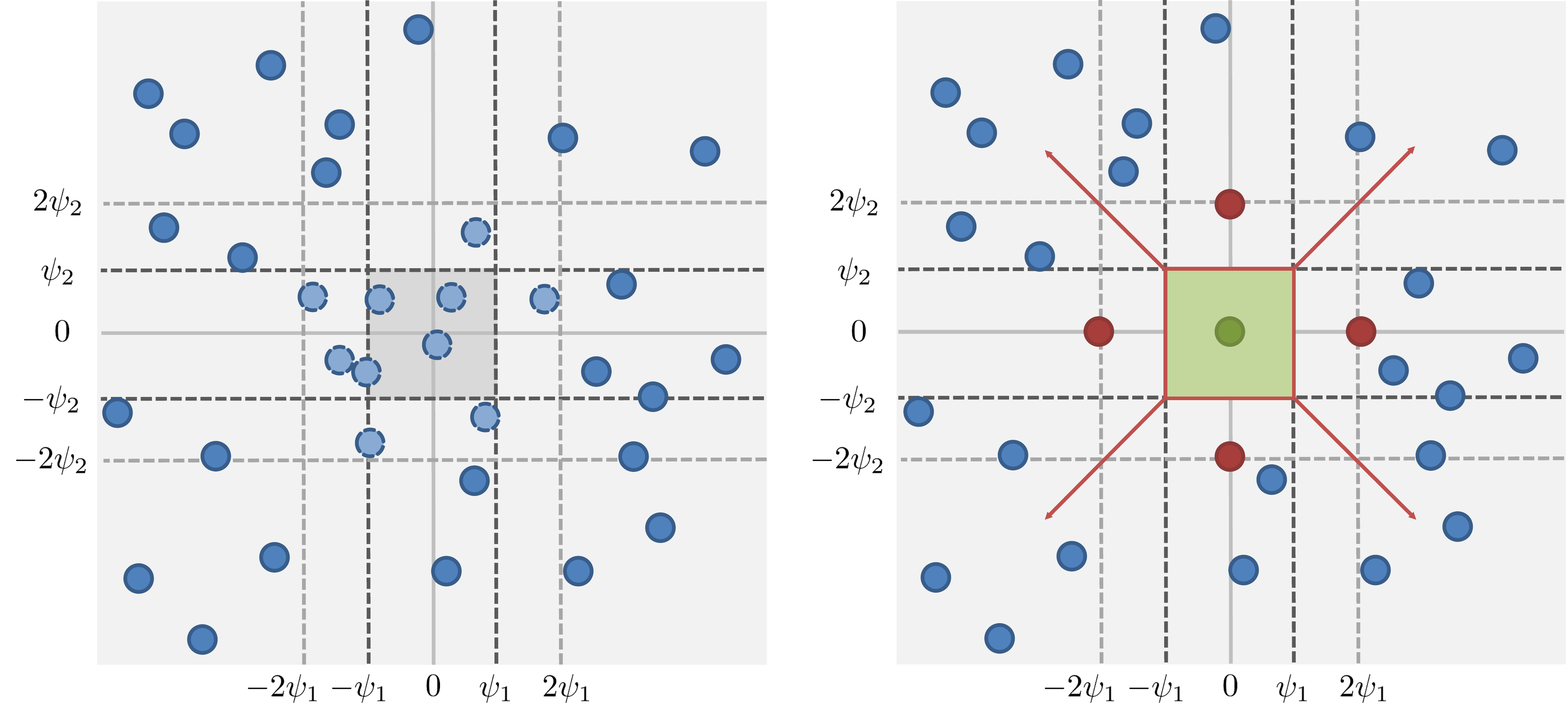}
  \caption{Preprocessing of \textit{k}-means input data to enforce the success state $\StateOpt$.
  \textbf{Left}: Continuous state-space with observed objective vectors \ObjectiveVec~(blue points). The points with dashed outline will be canceled out.
  \textbf{Right}: Delineation of $\StateOpt$ in green, enforced by inserted vectors (green / red points) with large weights. See text for details.}
  \label{fig:method:state_quantization_preprocessing}
\end{figure}

First, the $\vec{0}$-vector is inserted into $\ObjectiveSet$, along with two vectors per dimension $i$, where all components are zero, except the $i$\textsuperscript{th} component, which is set to $\pm 2 \Convergence_i$.
These $2 \nObjectives + 1$ inserted vectors are later converted into centroids of representative states to delineate the desired hyperrectangle for $\StateOpt$ as illustrated in \autoref{fig:method:state_quantization_preprocessing}.
Furthermore, to avoid malformation of $\StateOpt$, no other representative state should emerge within that region. 
Therefore, all vectors $\ObjectiveVec \in \ObjectiveSet$, where $\forall i: |\Objective_i| < 2 \Convergence_i$ (except for the inserted vectors) are canceled out by assigning zero weight, while the inserted vectors are assigned large weights $\to \infty$ and all remaining vectors weights of $1$.

Next, \textit{k}-means is initialized by placing a subset of the initial centroids at the locations of the inserted states, and the remaining $\nStates - 2 \nObjectives - 1$ centroids at random vectors in $\ObjectiveSet$.
Both the large weight and the custom initialization enforce the algorithm to converge to a solution where one cluster centroid is located at each inserted vector, while the other centroids are distributed according to the training data.
To ensure equal contribution of all objectives (cancel out different units, etc.), similarity is defined relative to the inverse of the user-defined convergence criteria (\autoref{eq:method:distance_measure}).

Finally, after \textit{k}-means converged, the resulting centroids, denoted $\CentroidVec_\State$, are used to delineate the region in $\mathbb{R}^\nObjectives$ assigned to a representative state \State.
%
%
%
\section{Data-driven Initialization}
\label{sec:appendix:initialization}
%
This section describes the details of the proposed data-driven initialization approach \modified{to compute a list of candidate initialization parameter vectors $\mathcal{X}_0 = (\ParamVec_0', \ParamVec_0'', \dots)$} for a new patient $p$ based on the patient's measurements $\MeasurementVec^p$ and the training database \EpisodeSet~(see \autoref{subsubsec:method:initialization}).

First, all model states are extracted from the training database: $\Upsilon = \{\ModelStateVec \in \EpisodeSet\}$.
Next, $\Upsilon$ is fed to a clustering algorithm (e.g.~\textit{k}-means).
As in \autoref{sec:appendix:quantization}, the distance measure is defined relative to the inverse of the convergence criteria (\autoref{eq:method:distance_measure}).
The output is a set of centroids (for simplicity, in this work the number of centroids was set to \nStates), and each vector is assigned to one cluster based on its closest centroid.
Let $\Upsilon_p \subseteq \Upsilon$ denote the members of the cluster whose centroid is closest to $\MeasurementVec_p$ and $\Xi_p = \{\ParamVec \in \EpisodeSet \mid \CompModel(\ParamVec) \in \Upsilon_p \}$ the set of corresponding model parameters.
For each cluster, an approximation of the likelihood over the generating parameters is computed in terms of a probability density function.
In this work a Gaussian mixture model is assumed:
\begin{equation}
  \text{GMM}_p(\ParamVec) = \sum_{m=1}^M \nu_m \mathcal{N}(\ParamVec; \boldsymbol{\mu}_m, \Sigma_m) \enspace .
  \label{eq:method:gmm}
\end{equation}
The parameter vectors in $\Xi_p$ are treated as random samples drawn from GMM$_p$.
Its properties, namely the number of mixture components $M$, their weights $\nu_m$, and their means $\boldsymbol{\mu}_m$ and covariance matrices $\Sigma_m$, are estimated from these samples using a multivariate kernel density estimator with automated kernel bandwidth estimation, see \cite{kristan2011multivariate} for more details.
Finally, the $M$ estimated means are selected as initialization candidates \modified{and stored in a list $\mathcal{X}_0 = (\boldsymbol{\mu}_{m'}, \boldsymbol{\mu}_{m''}, \dots)$.
The elements of $\mathcal{X}_0$ are sorted in descending order according to their corresponding $\nu_m$-values to prioritize more likely initializations: $\boldsymbol{\mu}_{m'}$ is the mean with $m' = \argmax_m{\nu_m}$.}
%
%
%
%
\section*{References}
\bibliographystyle{elsarticle-harv}
\bibliography{./biblio}
\end{document}